%% file: ElasticTreePaper.tex
\documentclass{sig-alternate}

\usepackage{epsf,epsfig}
\usepackage{amssymb,amsmath}
\usepackage{paralist}
\usepackage{url}
\usepackage{balance}

\newcommand{\iaas}{{\sl IaaS}}
\newcommand{\sla}{{\sl SLA}}
\newcommand{\slas}{{\sl SLA}s}
\newcommand{\udf}{\emph{UDF}}
\newcommand{\udfs}{\emph{UDF}s}
\newcommand{\vm}{{\sl VM}}
\newcommand{\vms}{{\sl VM}s}
\newcommand{\sysname}{{\sc Exareme}}

\def\sharedaffiliation{
\end{tabular}
\begin{tabular}{c}}

\def\stopproof{\square}
\def\square{\vbox{\hrule height.2pt\hbox{\vrule width.2pt height5pt \kern5pt
\vrule width.2pt} \hrule height.2pt}}

\begin{document}

\title{Elastic Processing of Analytical Query Workloads \\ on IaaS Clouds}

\numberofauthors{5}

\author{
\alignauthor Herald Kllapi$^*$
\alignauthor Panos Sakkos$^-$
\alignauthor  Alex Delis$^{*+}$
\and
\alignauthor Dimitrios Gunopulos$^*$
\alignauthor Yannis Ioannidis${^{*+}}$
\sharedaffiliation
\\
	\affaddr{*MaDgIK Lab, Dept. of Informatics and Telecommunications,} \\
	\affaddr{University of Athens, Athens 15784, Greece.} \\
      \affaddr{email: \{herald, ad, dg, yannis\}@di.uoa.gr} \\
	\affaddr{-Microsoft, Norway, email: panoss@microsoft.com,
work done at University of Athens} \\
	\affaddr{+``Athena'' Research Center, Athens 15125, Greece}
}

\maketitle

\input{abstract}

\input{introduction}
\input{motivation}
\input{notation}
\input{example}
\input{approach}
\input{experiments}
\input{related}
\input{conclusions}

\section*{Acknowledgments}
We thank Lefteris Stamatogiannakis and Marialena Kyriakidi for their helpful comments and Alexandros Papadopoulos for setting up the systems and running part of the experiments. This work is partially supported by the European Commission under contract 318338 (Optique project~\cite{book-big-data-2013}).

\balance

\begin{small}

\end{small}

\end{document}

%% file: abstract.tex
%!TEX root = ElasticTreePaper.tex

\begin{abstract}

Many modern applications require the evaluation of analytical queries on large amounts of data. Such queries entail joins and heavy aggregations that often include user-defined functions (\udfs). The most efficient way to process these specific type of queries is using tree execution plans. In this work, we develop an engine for analytical query processing and a suite of specialized techniques that collectively take advantage of the tree form of such plans. The engine executes these tree plans in an elastic \iaas\ cloud infrastructure and dynamically adapts by allocating and releasing pertinent resources based on the query workload monitored over a sliding time window. The engine offers its services for a fee according to service-level agreements (\slas) associated with the incoming queries; its management of cloud resources aims at maximizing the profit after removing the costs of using these resources. We have fully implemented our algorithms in the {\sysname} dataflow processing system. We present an extensive evaluation that demonstrates that our approach is very efficient (exhibiting fast response times), elastic (successfully adjusting the cloud resources it uses as the engine continually adapts to query workload changes), and profitable (approximating very well the maximum difference between \sla-based income and cloud-based expenses).

\end{abstract}

%% file: introduction.tex
%!TEX root = ElasticTreePaper.tex

\section{introduction}
\label{sec:intro}

\begin{figure}[t!]
\centerline{\psfig{figure=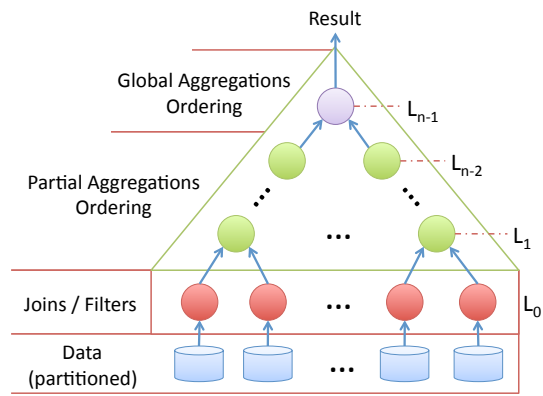,width=8cm} }
\vspace{-0.5cm}
\caption{Generic form of tree execution plans.}
\label{fig:tree-exec-plan}
\end{figure}

Many modern applications face the need to process voluminous data using ad-hoc analytical queries~\cite{saj, DBLP:journals/sp/PikeDGQ05, DBLP:conf/vldb/Simitsis03}. They also call for the use of complex user-defined functions (\udfs) that do not come from a pre-defined set of operators with well known semantics for which {\tt SQL} proper is often not sufficient or efficient to use.  Furthermore, these queries must demonstrate very fast and near-interactive response times~\cite{impala,
DBLP:journals/pvldb/AbrahamABBCGMMRSWZ13, DBLP:journals/pvldb/MelnikGLRSTV10}. It has been shown that, in appropriate computational environments such as shared-nothing, specific tree execution plans, can answer queries of the above kind on trillions of objects in seconds~\cite{DBLP:journals/pvldb/AbrahamABBCGMMRSWZ13, DBLP:journals/pvldb/MelnikGLRSTV10}.  Figure~\ref{fig:tree-exec-plan} shows a generic image of such a tree execution plan: the leaves of the tree represent the data that are partitioned appropriately based on the application. The remaining nodes represent operators (e.g., such as \verb group  \verb by s) and the connections between them correspond to operator dependencies. The operators at the first level ($L_0$) typically perform joins and filtering. The internal operators (levels $L_1$ to $L_{n-2}$) perform partial aggregations. Finally, the root operator (level $L_{n-1}$) performs global aggregations and produces the final result.

Several systems have been proposed for large-scale data processing~\cite{impala, DBLP:journals/pvldb/ChattopadhyayLLMALKW11, DBLP:journals/pvldb/MelnikGLRSTV10, DBLP:conf/icde/ThusooSJSCZALM10}; they are typically built on top of \iaas\ clouds~\cite{DBLP:journals/cacm/ArmbrustFGJKKLPRSZ10, DBLP:journals/ccr/GonzalezMCL09} which have emerged as an attractive platform for analytical query processing. The defining characteristic that favors \iaas\ clouds over other competing environments (such as distributed, cluster-based, grid, etc.) is \emph{elasticity}, i.e., the ability to lease compute and storage resources on--demand and use them only for as long as needed. This makes possible to create an \emph{elastic virtual infrastructure} that may change over time. \iaas\ clouds offer compute resources in the form of virtual machines (\vms). The cost of leasing a \vm\ is determined based on a per time-quantum pricing scheme, where one pays for the entire quantum independently of the extent of the use of the \vm\ resources~\cite{Amazon:WS}. An elastic cloud-enabled engine may allocate or de-allocate \vms\ dynamically, trying to identify the optimal trade-off between the need to minimize execution times for a given workload and the requirement to minimize the monetary cost of using the cloud resources~\cite{DBLP:journals/sigmod/FlorescuK09, DBLP:conf/sigmod/KllapiSTI11}.

\begin{figure}[t!]
\centerline{\psfig{figure=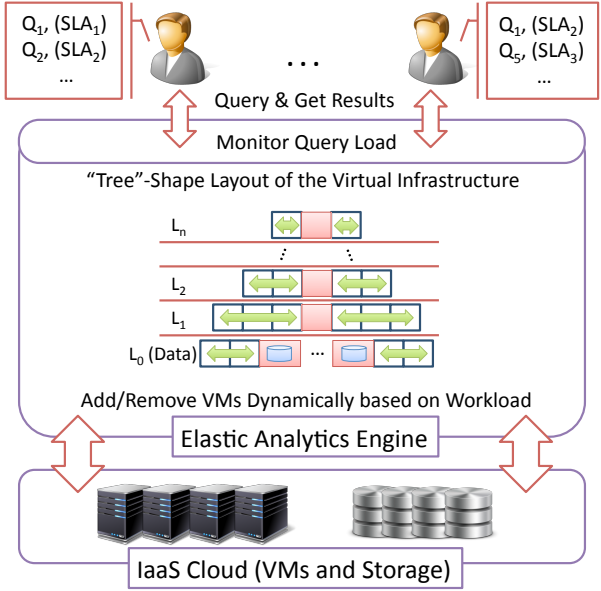,width=10.5cm} }
\vspace{-0.2cm}
\caption{Engine for Elastic Analytical Query Processing.}
\label{fig:system_setting}
\end{figure}

In this work, we develop an elastic processing engine operating atop an \iaas\ infrastructure that is capable of executing efficiently and cost-effectively a large class of analytical queries demonstrating a tree execution plan of a specific form. We have implemented the functionality within {\sysname}~\cite{DBLP:conf/owled/KllapiBHIJKKZ13, DBLP:journals/debu/TsangarisKKPPPSSI09}, 
our system for dataflow execution on the cloud. Figure~\ref{fig:system_setting} depicts the salient characteristics of our engine: arbitrarily complex queries, possibly having \udfs\  with arbitrary user--code, are continually submitted to the engine. Each query is associated with an \sla\  that designates the price that a query instigator must pay for answering the query depending on its response time (faster response times are associated with higher prices). The data is originally stored on the cloud 
(e.g., Amazon S3\cite{Amazon:WS}) and is partitioned 
to increase flexibility and performance.

In this context, our proposed engine and its requisite mechanisms make the following contributions:

\begin{itemize}
\itemsep 0pt

\item  We introduce an online algorithm that exploits the elasticity of \iaas\ clouds to {\em adapt} the size of the virtual infrastructure to the query workload at hand by dynamically allocating or de-allocating \vms. This is done so that our engine maximizes its profit while taking into account the monetary cost of expended cloud resources as well as the \slas\ of the submitted queries.

\item We propose to lay out the \vms\ allocated in a ``tree'' shape (Figure~\ref{fig:system_setting}), so that query execution plans are mapped naturally to \iaas\ processing elements. The \vms\ at the leaf-level \emph{fetch data} from the cloud storage and cache it to their local (virtual) disk for processing, thereby decoupling compute and storage resources. For partition assignments, we use an extension of \emph{consistent hashing} and devise a simple, yet quite accurate, analytical formula to approximate the cost of partition reassignment; we use this formula when our online algorithm searches for an optimal choice when considering changes in the deployment of resources at the data level $L_{0}$ (as shown in Figure~\ref{fig:system_setting}).

\item We have implemented our approach within {\sysname} and have performed an extensive experimental evaluation which indicate significant and very promising results. Our method compares favorably to \emph{Cloudera Impala}~\cite{impala} on sheer performance offering near-interactive response times, it adapts quickly to workload changes, and it increases the processing engine profit significantly compared to static infrastructures.

\end{itemize}

The rest of this paper is organized as follows: Section~\ref{sec:motivation} offers motivating query examples from two key classes of contemporary query processing and Section~\ref{sec:notation} discusses the operating environment. Section~\ref{sec:setting} outlines the intuition for our suggested solution and Section~\ref{sec:algs} presents the proposed query engine. Section~\ref{sec:exps} furnishes our key experimental findings while related work and conclusion are found in Sections~\ref{sec:related} and~\ref{sec:future} respectively.

%% file: motivation.tex
%!TEX root = ElasticTreePaper.tex

\section{Motivation - Tree Queries}
\label{sec:motivation}

We draw our motivation from key classes of analytic queries frequently encountered in data warehouses and {\sl NoSQL}-systems.

\noindent
{\em i) Data Warehouses} store historical data used to help understand market trends and create management reports~\cite{DBLP:conf/icde/ThusooSJSCZALM10}. Typical queries perform joins and extensive aggregations, and usually return only heavy hitters (the top records as ordered on some columns)~\cite{DBLP:conf/damon/PolychroniouR13}. The following query shows such an example in {\tt SQL} that is inspired by the {\sl TPC-H} benchmark~\cite{tpch}:

\begin{small}
\begin{verbatim}
SELECT year, country,
       sum(l_extendedprice) as revenue,
       SUMMARY(l_extendedprice) as report
FROM lineitem, supplier, nation
WHERE l_suppkey = s_suppkey
  AND s_nationkey = n_nationkey
GROUP BY year, country ORDER BY year, country;
\end{verbatim}
\end{small}

The query joins three tables, groups the results by each country and year, and computes the revenue for each group. It also uses the \verb SUMMARY   \udf\ to generate a report on the overall output.

The typical schema of a data warehouse is a star or a snowflake~\cite{DBLP:reference/db/MorfoniosI09a} and is heavily denormalized for performance. The fact table \verb lineitem   in the example, is very large compared to other two tables. To expedite processing, the data placement here has the fact table partitioned horizontally and the other tables replicated at all locations where partitions exist. Thus, all query joins are local to each machine and the aggregations can be executed as a tree.

\noindent
{\em ii) {\sl NoSQL}--systems} provide techniques to store and process data that is typically in the form of key-value pairs, graphs, or documents~\cite{DBLP:conf/osdi/DeanG04, DBLP:journals/pvldb/LowGKBGH12, DBLP:conf/sigmod/OlstonRSKT08}. Typical queries involve filtering and transformations on a single input table while joins are usually avoided as they are often expensive; required joins can be realized atop such systems~\cite{DBLP:conf/sigmod/OkcanR11}. The following example dataflow shows how a simple intrusion detection analysis on server logs could be expressed in \emph{FlumeJava}~\cite{Chambers:2010}:

\begin{small}
\begin{verbatim}
PCollection<String> in = ReadInput("log.txt");
// Parse and convert to log entry objects
PCollection<KV<IP, LogEntry>> entries =
    in.parallelDo(new LogTransform());
PTable<IP, Collection<LogEntry>> g =
    entries.groupByKey();
// Perform analysis on each group g
PTable<IP, Report> result =
    g.combineValues(new IntrusionAnalysis());
FlumeJava.run();
\end{verbatim}
\end{small}

\noindent
The dataflow reads the input from file \verb log.txt  (one row per line) and converts it to key-value pairs using the \verb LogTransform  \udf\ with the respective IP as key. It then groups entries by IP and performs an intrusion detection analysis on each group using the \verb IntrusionAnalysis  \udf. The usual data placement has the files partitioned in blocks of fixed size and distributed to different \vms, typically using a distributed file system~\cite{DBLP:conf/sosp/GhemawatGL03}. In this example, \verb LogTransform  is executed in parallel on all blocks of the file and \verb IntrusionAnalysis  is executed again in parallel on the formed groups.

The solutions proposed in the past for the above categories 
of queries are not sufficient for cloud environments for they 
 \begin{inparaenum}[\itshape a\upshape)]
\item
 treat all resources indistinguishably with no attention to the nature of the queries, 
\item
are not elastic, and/or
\item they target performance by evaluating queries as fast as possible, treating the monetary cost as a secondary consideration or ignoring it completely. 
\end{inparaenum}
Our work comes to fill this gap.

%% file: notation.tex
%!TEX root = ElasticTreePaper.tex

\section{Problem Formulation}
\label{sec:notation}

We present in details key aspects of the problem we address together with the relevant notation and definitions.

\subsection{IaaS Cloud}
A {\em container} or \vm\ is the unit of cloud {\bf compute resources} and includes CPU(s), memory, disk(s), and network resources. All containers furnished for general use have the same size, i.e., the same capacity in every type of resource they provide, e.g., equal memory size. By and large, this is typical of most clouds  where only a limited number of \vms\ has substantially enhanced resources to help them run core services (e.g., \emph{namenodes} for \emph{Hadoop}~\cite{apache:hadoop}). The price $M^c_Q$ for using a container is a fixed amount in \$ per time quantum $T_Q$. The set of containers allocated to a cloud application, such as our query processing engine, constitutes the  {\bf virtual infrastructure} of the application. The cloud also offers {\bf data storage resources}, which are decoupled from its compute resources for flexibility. \vms\ transfer data from these storage resources and cache it to their local virtual disks for processing.

\subsection{Data Partitioning}
Tables are {\bf partitioned} and {\bf replicated} so that joins (if any) are local to containers and only aggregations require data transfer. Hence, partitioning is based on foreign keys used in joins. If the database has only one table (the usual case in {\sl NoSQL}--systems), it is partitioned randomly into shards of equal size. If the database has multiple tables as it happens in data warehouses, the largest tables (one or more, depending on the available storage) are partitioned and all others are replicated wherever the partitions are stored. In this regard, in the {\sl TPC-H} benchmark, it may be most beneficial to partition the two largest tables \verb lineitem  and \verb orders  with hash partitioning on \verb l_orderkey , which is a foreign key in table \verb orders , and replicate the other tables. This is precisely the partitioning scheme we use for {\sl TPC-H} in our experiments.

\subsection{Properties of Analytical Queries}
Issued {\tt SQL} queries may include filters, joins, and two types of group aggregate functions: {\bf distributive} and {\bf algebraic}~\cite{DBLP:books/mk/HanK2000}. Distributive functions are directly parallelizable, as they are commutative, associative, and for a table $T$ with two partitions $T_1$, $T_2$, satisfy the property $f(T)=f(f(T_1)\cup f(T_2))$. Examples of such functions from {\tt SQL} include \verb min , \verb max ,  and \verb sum. Algebraic functions are indirectly parallelizable, as they can be expressed as algebraic combinations of distributive or other algebraic functions. Examples from {\tt SQL} include \verb count , \verb avg , \verb stdev , all expressed as increasingly more complex combinations of \verb count  and \verb sum . More importantly, the queries we support may also include \udfs\ with arbitrary code that may correspond to distributive or algebraic functions. A \udf-example is the function of \verb reservoir  \verb sampling ~\cite{DBLP:journals/toms/Vitter85} which randomly selects a subset of a table's records with equal probability.

Using the above properties, we may readily transform flat queries into tree plans by recursively unwrapping all algebraic functions until only distributive functions are left. For example, consider two tables \verb R(A,B, ...)  and \verb S(B, ...), both partitioned on column \verb B , and the following flat query:

\begin{small}
\begin{verbatim}
select avg(A) as AA from R, S where R.B = S.B
\end{verbatim}
\end{small}

We transform the above {\tt SQL}-statement into a tree-based one using the following four ``conceptual queries'': {\em leaf}, {\em internal-initial}, {\em internal-recursive}, and {\em root}. The particulars of each query are as follows:

$\bullet$ {\bf Leaf}: carrying out filtering and joins 
\begin{small}
\begin{verbatim}
select A from R, S where R.B = S.B;
\end{verbatim}
\end{small}

$\bullet$ {\bf Internal-initial}: executing the distributive aggregate initialization 
\begin{small}
\begin{verbatim}
select sum(A) as SA, count(*) as CA from leaf;
\end{verbatim}
\end{small}

$\bullet$ {\bf Internal-recursive}: producing partial distributive aggregation(s)
\begin{small}
\begin{verbatim}
select sum(SA) as SA, sum(CA) as CA
from internal-initial;
\end{verbatim}
\end{small}

\noindent
$\bullet$ {\bf Root}: compiling sought algebraic aggregation(s)
\begin{small}
\begin{verbatim}
select sum(SA) / sum(CA) as AA
from internal-recursive;
\end{verbatim}
\end{small}

The above conceptual queries have to be placed on the morphed query execution tree (e.g., Figure~\ref{fig:tree-exec-plan}). The {\em leaf} queries are placed at level $0$ of the execution tree in order to be executed in parallel on each partition. Since {\em internal-initial} also functions on each partition independently, this type of query can be part of level $0$. Between level $1$ of the tree (e.g., Figure~\ref{fig:tree-exec-plan}) and its root, we place {\em internal-recursive} queries. Given the commutativity and associativity of distributive functions, there may be an arbitrary number of levels of {\em internal-recursive} queries, without affecting correctness. The actual number of the internal level of the resulting query--tree depends on the size of the original tables and the affordable degree of parallelization. Finally, note that, for a query without algebraic functions, the {\em root} query is identical to the {\em internal-recursive} query.

\subsection{Service Level Agreement}
An \sla\ is a function having query execution time as input and money as output, namely, $SLA: \mathbb{R}^+ \rightarrow \mathbb{R}$, both in appropriate units, often in  seconds and dollars respectively. \slas\ can be step-wise or more sophisticated~\cite{xiong2014admission, DBLP:conf/icde/TsakalozosKSRPD11}. Inspired by other works, we use a generic form of \slas\ defined as follows: $SLA(u, q, t) = \alpha \cdot e^{-t/\gamma}$, where $\alpha$ and $\gamma$ are respectively regulators of the maximum amount of money a user pays and the monetary cost reduction rate with time. A small $\gamma$ indicates a critical query that should be rapidly executed as its value drops drastically. Alternatively, a large $\gamma$ indicates a best-effort query. We avoid using a step function because smoothness plays an important role in our optimization problem as we describe in Section~\ref{sec:algs}.

\begin{figure}[t!]
\centerline{\psfig{figure=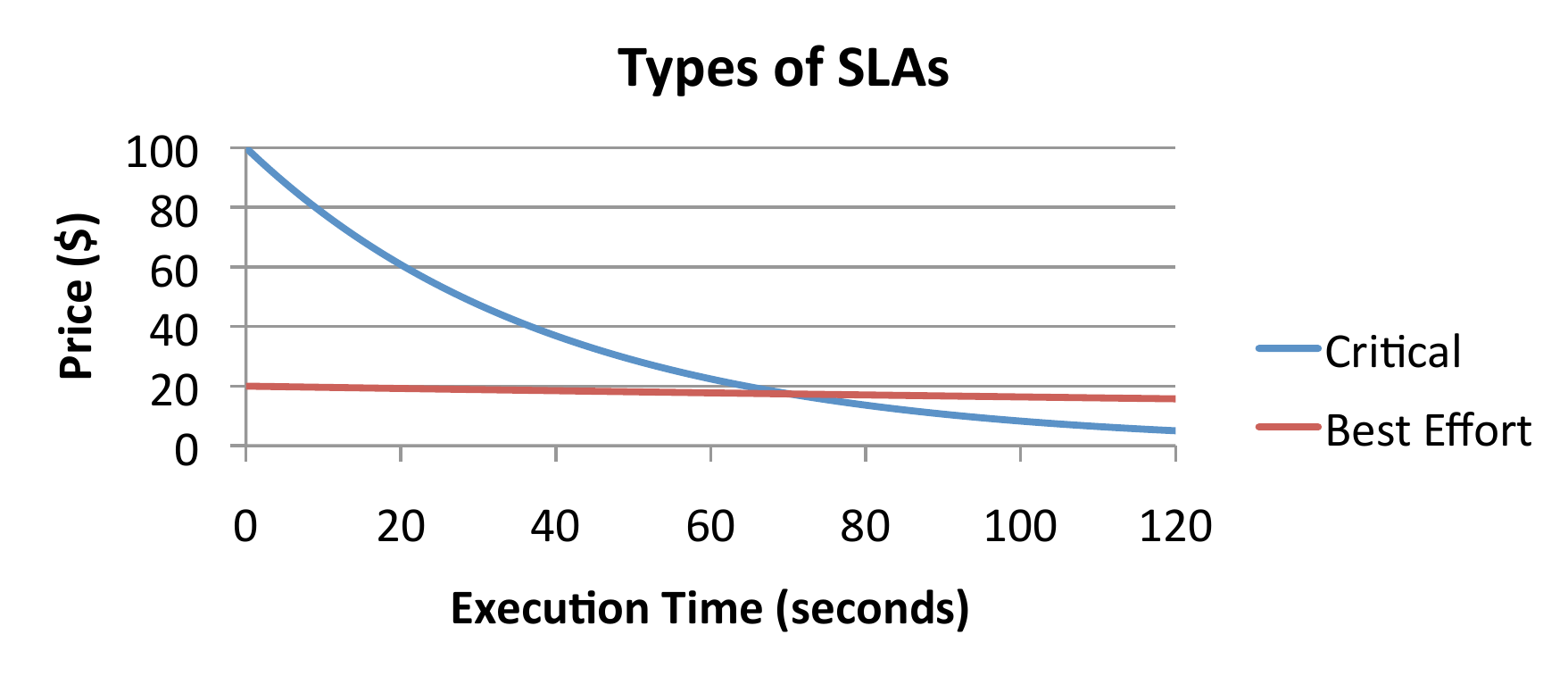,width=8.5cm}}
\vspace{-0.4cm}
\caption{Two SLAs: `critical' and `best-effort'.}
\label{fig:sla}
\end{figure}

An example of two different SLAs is shown in Figure~\ref{fig:sla}. The {\em critical} SLA has $\alpha=100$ \& $\gamma = 40$ and the {\em best-effort} has $\alpha=20$ \& $\gamma = 500$. Notice that the {\em critical} SLA is very profitable for low execution times but its price drops quickly. The definition of SLAs can be extended to include negative values: a penalty that the service provider pays if the execution time is large. We leave the exploration of this alternative as future work.

\subsection{Profit Maximization Problem}
The queries are issued to the engine in a streaming fashion. Each query is associated with its own \sla. The price of the query charged is computed using both its \sla\ and its execution time. The {\bf revenue} generated by the engine during a particular time period $p$ is computed as the summation of the prices all queries launched during the period in question. The {\bf operational cost} in $p$ using $c$ containers is computed as: $O= c \cdot p / M^{c}_{Q}$. The {\bf profit} $P$ during the same period is computed as $P = R-O$. Our optimization objective is to maximize the provider's profit during the operation of the engine, i.e., maximize the difference between operational cost and revenue.

%% The following figure is from the example section!
\begin{figure*} [t!]
\centering
\includegraphics[width=18.3cm]{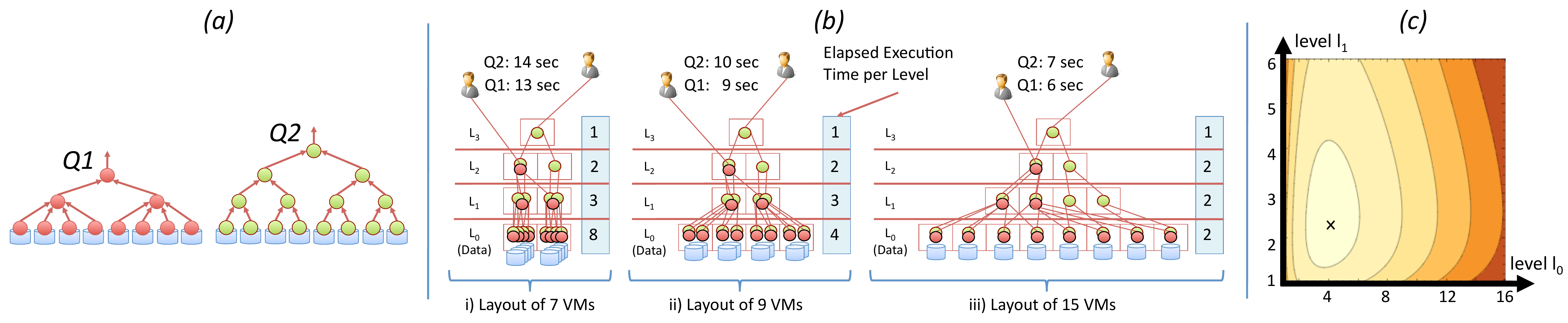}
\vspace{-0.4cm}
\caption{a) Execution plans for {\sl Q1/Q2}, 
	b) Mapping of the $2$ plans on $3$ different \vm\ layouts, 
	and c) Profit as a function of the number of \vms\ $l_0$ and $l_1$.
	}
\label{fig:example}
\end{figure*}

\begin{figure}[t]
\vspace{-0.7cm}
\centerline{\psfig{figure=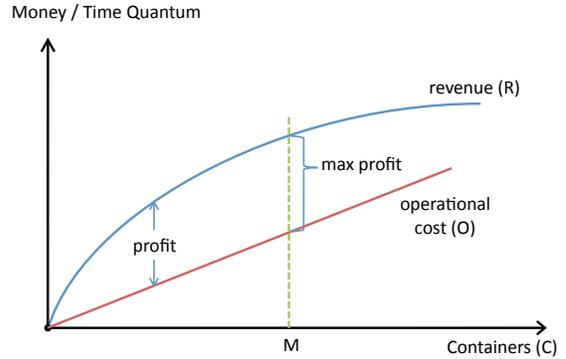,width=8.5cm} }
\vspace{-0.9cm}
\caption{Profit maximization based on revenue and operational cost.}
\label{fig:profit}
\end{figure}

Figure~\ref{fig:profit} illustrates our optimization goal; it shows a typical revenue curve per time quantum as affected by the number of containers~\cite{DBLP:conf/icde/TsakalozosKSRPD11}. The $y$-axis indicates the rate with which the revenue  is generated. The figure also shows the operational cost of the engine per time quantum, which is linear to the number of containers allocated as the incurred expense for every \vm\ by the provider is the same. Our goal is to identify a point $M$, that is the optimal number of containers that help maximize profit, i.e., the difference between revenue and operational cost is maximized. Notice that that the revenue function is a ``moving target'' as it highly depends on the query workload and so,  $M$ does \emph{change} over time. The engine should be able to dynamically adapt to workload changes and find the optimal point of operation at any moment.

%% file: example.tex
%!TEX root = ElasticTreePaper.tex

\section{Illustrative Example}
\label{sec:setting}

In this section, we present an example to give a high level overview 
of our approach. Figure~\ref{fig:example}(a) depicts two queries that are issued concurrently to the engine. Each query is transformed into a tree execution plan with its data at the leaves of the tree and respective operators at the internal nodes. For simplicity, assume that the execution time of each operator is $1$ second and that it generates some amount of data that is negligible. Further, assume that the \slas\ for both queries are identical and defined as: $price(t)=15\cdot e^{-t/20}$, where $t$ is the query execution time measured in seconds and the \emph{price} is measured in \$.

The engine has allocated several \vms\ from the cloud and the data is appropriately partitioned. We lay out the deployed \vms\ in a ``tree'' shape, to naturally map the execution plans of both discussed queries onto the allotted virtual infrastructure. This tree-shaped use of resources may lead to diverse deployments as Figure~\ref{fig:example}(b) illustrates; here, we depict three different execution \vm\ layouts that help materialize {\sl Q1} and {\sl Q2}. More specifically, layout \emph{(ii)} of Figure~\ref{fig:example}(b) works with $9$ \vms\ of which $4$ are at the data level ($L_{0}$), 2 \vms\ at each intermediate levels ($L_1$--$L_2$),  and $1$ \vm\ at the root ($L_3$).

The three different layouts of Figure~\ref{fig:example} render different processing times when concurrently executing {\sl Q1} and {\sl Q2}. For example in layout \emph{(ii)}, {\sl Q1} and {\sl Q2} complete at $9$th and $10$th seconds respectively. The turnaround times are computed by summing the delay each query faces at each level of the layout. Given that we have $16$ operators at $L_0$ and each one runs for $1$ second, the total delay using $4$ \vms\ is $4$ seconds. At $L_{1}$, we have $6$ operators ($2$ from {\sl Q1} and $4$ from {\sl Q2}) that yield a delay of $3$ seconds since $2$ \vms\ are used. Similarly at levels $L_{2}$ and $L_{3}$ the respective delays stand at $2$ and $1$ second.

\begin{table}[h!]
\centering
    \begin{small}
    \begin{tabular}{|l|r|r|r|}
    \hline
    {\bfseries} & {\bfseries layout A} & {\bfseries layout B} & {\bfseries layout C} \\
    \hline
    {\sl Q1} Time (sec) & 13 & 9 & 6 \\
    \hline
    {\sl Q1} Price (\$) & 7.83 & 9.57 & 11.11 \\
    \hline
    \hline
    {\sl Q2} Time (sec) & 14 & 10 & 7 \\
    \hline
    {\sl Q2} Price (\$) & 7.44 & 9.09 & 10.57 \\
    \hline
    \hline
    Revenue (\$) & 15.27 & 18.66 & 21.68 \\
    \hline
    \vm\ Cost (\$) & 7.00 & 9.00 & 15.00 \\
    \hline
    Profit (\$) & 8.27 & 9.66 & 6.68 \\
    \hline
    \end{tabular}
    \end{small}
\caption{Profit for the Different Three Layouts.}
\label{tab:infra_profit}
\end{table}

Assuming that the costs of each \vm\ is \$1 for simplicity, using the above execution times for each level, the \emph{price} formula computes the charged price for each query. Table~\ref{tab:infra_profit} shows both revenue (i.e., sum of prices) and profit made on the provided service. The latter is computed as the difference \emph{revenue-cost} and yields layout \emph{(ii)} as the best of the three choices in Figure~\ref{fig:example}(b).

If we are to automate the above procedure, we need to articulate execution times in diversified layouts given a number of \vms\ $l_i$ at each level $L_i$. Provided that the number of operators drops exponentially from the leaves to the root of the query tree, the two lower levels $L_0-L_1$ have the greatest impact as far as the turnaround time of queries is concerned. Should we assume that the \vm\ numbers $l_2$ and $l_3$ do not change, the execution times for {\sl Q1} and {\sl Q2} required for these two levels are $2$ and $3$ seconds respectively. The potential profit generated by levels $L_0-L_1$  when different numbers of \vms\ are deployed to materialize the two queries is as follows:

\begin{equation*}
\begin{split}
profit(l_0, l_1) = price(t_{Q1})+price(t_{Q2})-cost(l_0 + l_1)= \\
15 e^{(-t_{Q1} / 20)} + 15 e^{(-t_{Q2} / 20)} - (l_0 + l_1) = \\
15e^{(-(t(l_0,l_1)+3)/20)}+15e^{(-(t(l_0,l_1)+2)/20)} - (l_0+l_1)
\end{split}
\end{equation*}

where $t(l_0,l_1)=16/l_0+6/l_1$ is the time required for the concurrent execution of the two queries at levels $L_0-L_1$; here, $16$ is the total time needed by all operators at  $L_0$ carried out by $l_0$ \vms\ (assuming perfect load-balancing) and $6$ is the total time required by operators at $L_1$ which are ultimately carried out by $l_1$ \vms. Figure~\ref{fig:example}(c) plots the contour of the expected $profit()$ as a function of $l_0$ and $l_1$; the profit increases as we move from darker to lighter color. For this specific version of the layout problem, the contour plot points out that the optimal solution is around $4$ \vms\ at $L_0$ and $3$ \vms\ at $L_1$.

We have to generalize the solution for layout selection discussed above when we consider multiple parameters including the $l_i$ numbers of \vms\ allotted to every level, number of queries considered together, potential data re-organization, \slas\ as well as timing aspects of the engine's operation. In doing so, the following challenges arise:
\begin{inparaenum}[\itshape A\upshape)]
\item how to find the optimal number of \vms\ given a query workload, 
\item how to schedule the query execution trees on the available \vms, 
\item how to {\em dynamically change} the layout and adapt to changes in the workload, and 
\item how to partition the data in order to add or remove \vms\  without significant network overhead. 
\end{inparaenum}
We address each of these challenges in the following sections.

%% file: approach.tex
%!TEX root = ElasticTreePaper.tex

\section{Overall Approach}
\label{sec:algs}
In this section, we present the overall approach we use to maximize profit. Time is separated into windows of fixed length (e.g., epochs of $300$ seconds) and inside each window we do not adjust the virtual infrastructure. All queries issued within a window, are scheduled assuming a fixed container layout. In the beginning of each window, we compute the new layout based on the measurements collected from the queries in a number of previous time windows while taking into account data re-configuration cost. In this section, we discuss the data partitioning scheme we employ, the elastic container layout, our online elastic layout allocation approach, and the query scheduler we use.

\subsection{Container Layout}
A {\bf container layout} is a hierarchical overlay on top of the allocated containers that defines the allowed communication channels between them. Figure~\ref{fig:system_setting} shows this generic layout. Each level has a fixed number of initial containers (shown in red in the figure) and is elastic, i.e., can change in size by allocating or deleting containers while enforcing optional minimum/maximum thresholds. The table partitions are located at the lowest level of the layout. Each \vm\ found at internal level $L_i$ can communicate {\em only with} the levels above ($L_{i+1}$) and below ($L_{i-1}$). Trees with height of $4$ or more are rarely needed in practice and only appear in very large data centers~\cite{DBLP:journals/pvldb/MelnikGLRSTV10}. For this reason, we use $3$ levels in our setting, however this is configurable.

\subsection{Data Partitioning and Placement}
\begin{figure}[t!]
\centerline{\psfig{figure=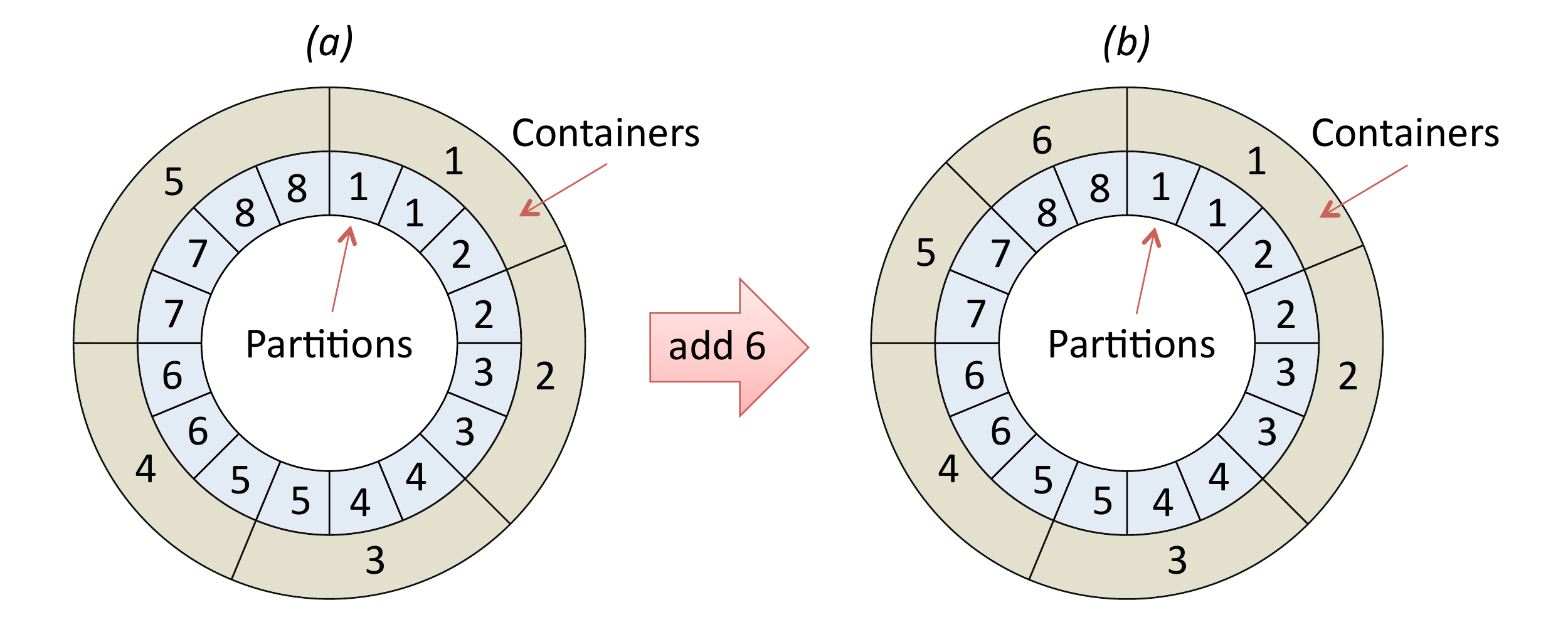,width=8.5cm} }
\vspace{-0.2cm}
\caption{Partitioning and Placement of Data using consistent hashing.}
\label{fig:consistent-hash}
\end{figure}

Our method is based on \emph{consistent hashing (CH)}~\cite{DBLP:conf/stoc/KargerLLPLL97} as its present good theoretical bounds on the size of data required to move when containers are added or deleted. Table partitions are placed in a logical circle as shown in the inner-circle of Figure~\ref{fig:consistent-hash}(a). The outer circle consists of the deployed containers at $L_0$ with each one assigned one or more partitions. For example, partition $3$ is assigned to container $2$. Notice that we place each partition multiple times in the inner circle. The first time a partition is accessed from the cloud storage is cached for subsequent usage. When a new container is added, it is placed in the outer circle at a position next to the container having the largest number of partitions; the latter sheds half of its data partitions to the new arrival. For example, when new container \#6 is added, is placed next to \#5; Containers \#5 and \#6 then split the existing partitions as shown Figure~\ref{fig:consistent-hash}(b).

To increase parallelism and flexibility we use {\bf over-partitioning} and {\bf replication}. We partition the tables into many more parts than the number of maximum data containers predicted to use (e.g., $10$ times more). Thus, changing the number of containers will cause only data transfers between the cloud storage and \vms, yet, it does not call for extensive re-partitioning (e.g., using hashing) on the cloud storage; this last operation is in general very expensive and incurs high network traffic~\cite{DBLP:journals/pvldb/LowGKBGH12}. Furthermore, we employ replication by adding each partition multiple times to the inner circle of Figure~\ref{fig:consistent-hash} in adjacent positions. Thus, when high parallelism is needed, the same partition will be assigned to multiple containers. Here, we balance the load between the containers that are assigned replicas of a partition. If more than one replicas happen to be assigned to the same container, we keep only one copy.

Load balance is crucial in our setting since the execution time of the operators at each level of the layout is bounded by the operator with the maximum execution time. Partitioning skew will delay the execution of all queries and affect revenue. For this reason, we extend the baseline--\emph{CH} to make it ``more aggressive'' when adding or removing containers as follows: instead of splitting partitions between two containers, we perform a local balancing around the insertion point and split partitions among the nodes in vicinity of \emph{Arc}+$1$ containers. This way, the re-organization is still local in the circle but the partitioning is more balanced. In practice, we use a window of size \emph{Arch}=$4$.

\begin{figure}[t!]
\vspace{-0.1cm}
\centerline{\psfig{figure=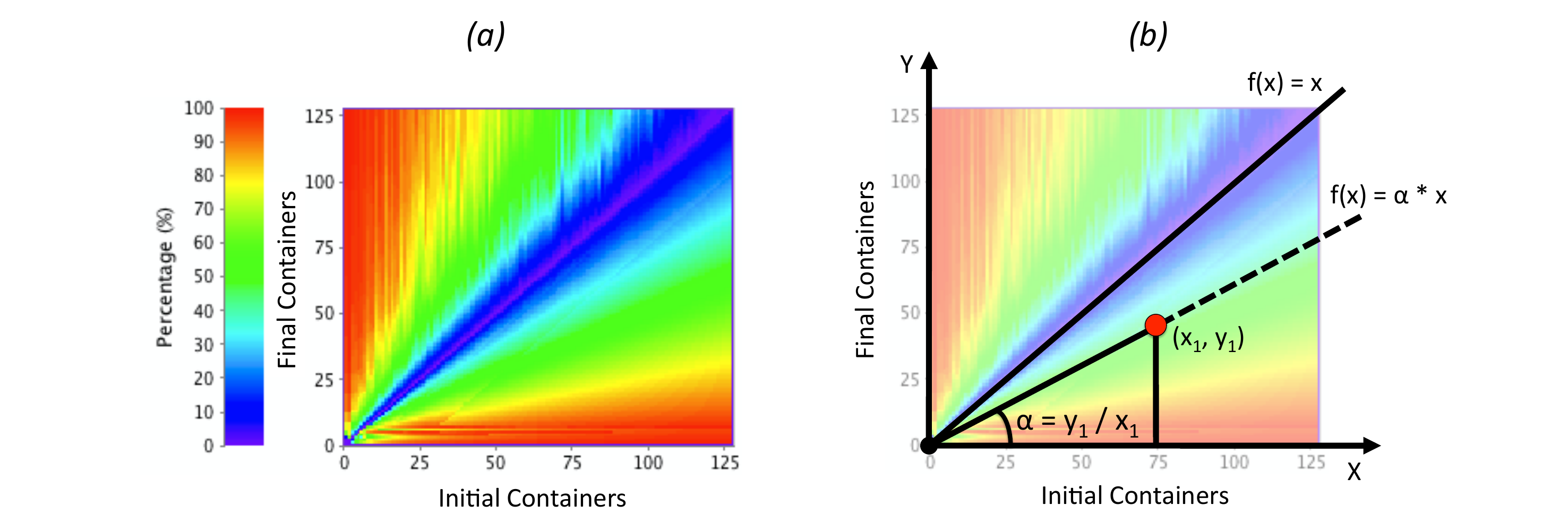,width=11.0cm} }
\vspace{-0.4cm}
\caption{Percentage of partitions assigned to a different container when changing their number (a) and the modeling of data movement (b).}
\label{fig:data-move-model}
\end{figure}

Figure~\ref{fig:data-move-model}(a) presents the outcome of an experiment using \emph{CH} with $128$ partitions and replication degree $3$ whose goal is to demonstrate the robustness of the method. The $x$ and $y$ axes show the initial and final number of containers (i.e., going from $x$ to $y$ containers). If $x$$<$$y$, then new containers are allocated, otherwise are deleted. We observe that when the changes are near the diagonal of the $2D$--space, \emph{CH} is robust to changes as the percentage of partitions requiring for re-assignment remains low ($\leq$10\%). This characteristic makes \emph{CH} ideal as a partition placement policy for our elastic processing engine.

\begin{figure}[t!]
\centerline{\psfig{figure=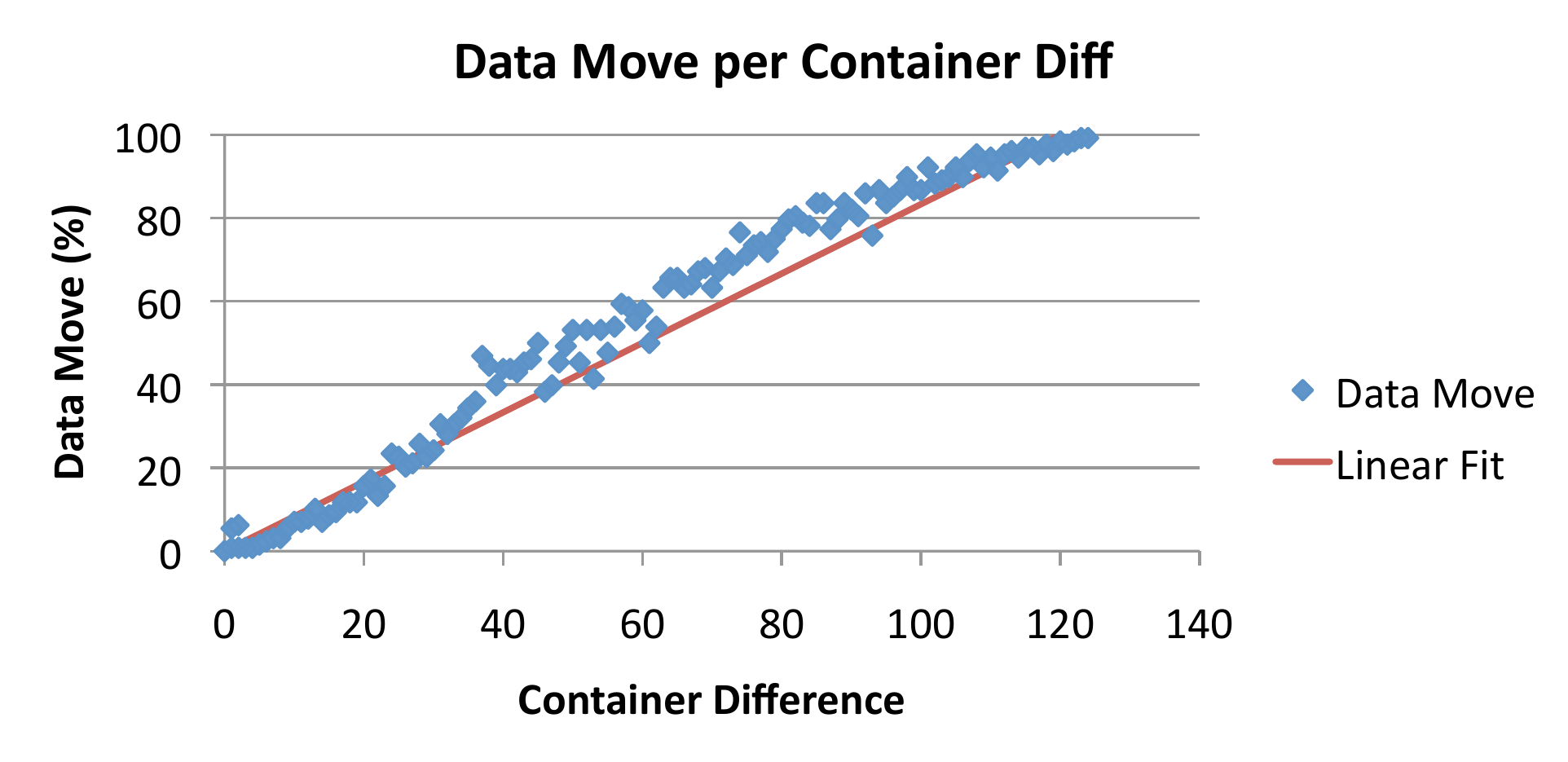,width=8.5cm} }
\vspace{-0.4cm}
\caption{A 1-D cut of Figure~\ref{fig:data-move-model}(a) at 125 containers.}
\label{fig:data-partitioning-cut}
\end{figure}

We need to model the above behavior of \emph{CH} to use it in our optimization process and thus, take into account data re-organization when adjusting the size of the deployed virtual infrastructure. Figure~\ref{fig:data-partitioning-cut} shows a $1D$ cut of the two dimensional plot of Figure~\ref{fig:data-move-model}(a) at 125 containers which reveals a strong linear correlation between the number of containers and the percentage of partitions moved. Figure~\ref{fig:data-move-model}(b) provides the sought model that predicts the data needed to be transferred when the number of \vms\ changes. Let $x$ and $y$ be the previous and new number of containers. The size of data that have  to move is modeled as:
$size_d(x, y) = (1 - min(x/y, y/x)) \cdot data\_size$, 
with $data\_size$ is the total volume of the tables taking into account partitioning and replication. Factor $min(x/y, y/x)$ is used to remove the symmetry of the $2D$--space on the diagonal for $size_d(x, y) = size_d(y, x)$. We measured the modeling error by computing the difference between the actual number of partitions moved (as shown in Figure~\ref{fig:data-move-model}(a)) and the predictions of our model and found that the estimation error to be on the average $6.4$\%, which is deemed very robust.

\subsection{Elastic Layout Allocation}
Our suggested algorithm for {\em Elastic Layout Allocation} helps dynamically change the container layout based on the query workload received to maximize profit. The proposed online algorithm works as follows: it uses the queries issued on a historical window $W_H$, their CPU load, and the data the queries transferred through the network. Using these statistics, the algorithm makes predictions for a window of size $W_P$ in the future~\cite{DBLP:journals/tcst/AngCL05, DBLP:conf/sigmod/DugganS14}. We model the profit as a multivariable function, representing each level of the container layout with a variable that indicates the number of containers allocated ($l_i$). The goal is to find the optimal number of containers in each level that maximize profit in the prediction window. In our experiments, we use a historical window of 2 epochs (i.e., $600$ seconds) to make predictions for the upcoming window of $300$ seconds. Notice that a large $W_H$ will cause the engine to adapt slowly to the workload and low $W_H$ may cause it to change rapidly: both extremes are not ideal. We experimentally ascertained that these window sizes behave well and leave for future work the automated learning of these numbers. Next we formally define our optimization function.

The queries are separated into a finite number of classes each having its own \sla\ which is the usual case in practice~\cite{DBLP:conf/icde/TsakalozosKSRPD11}. We denote as $\overrightarrow{\alpha}$ and $\overrightarrow{\gamma}$ the vectors carrying the respective values for all \slas. Let $\overrightarrow{Q_H}$ be the vector with the number of queries per \sla\  that have been executed during the historical window $W_H$. The total number of queries is $numQ_H = \sum_i(\overrightarrow{Q_H}[i])$. We denote as $\overrightarrow{L_H}$ the current number of containers allocated at each level of the layout. Similarly, $\overrightarrow{CPU_H}$ is the vector with the sum of CPU loads at every level of the layout within the historical window and $\overrightarrow{NET_{H}}$ is the total amount of data transferred outwards every level. Furthermore, we designate $conc$ to be the average number  of queries running concurrently at any point in time. We compute $conc$ by summing the execution times of all queries within the historical window and divide this number by the length of this window. All the concurrently running queries share the same resources, and thus, they implicitly affect each other.

Dealing with the prediction window $W_P$, we denote as $\overrightarrow{L_P}$ the container topology computed. Using the historical measurements and $ \overrightarrow{L_P}$, we can predict the average running time of the queries in the prediction window as follows:
\begin{equation*}
\begin{split}
t_P = \frac{conc}{numQ_H} \Big[ \frac{\overrightarrow{CPU_{H}}[1]}{\overrightarrow{L_P}[1]} + \\
\sum_{i=2}^{|\overrightarrow{L_P}|} \Big(\frac{\overrightarrow{CPU_{H}}[i]}{\overrightarrow{L_P}[i]} +
\frac{\overrightarrow{NET_H}[i]}{net\_speed \cdot min(\overrightarrow{L_P}[i-1], \overrightarrow{L_P}[i])} \Big) \Big]
\end{split}
\end{equation*}
where $\overrightarrow{CPU_{H}}[i] / \overrightarrow{L_P}[i]$ is the CPU load per container at level $i$ of the layout. The factor $1 / numQ_H$ above calculates the average time expended per query and we have to multiply by $conc$ in order model the delay that each query poses on others running concurrently. At this point, our model assumes that it can achieve perfect load-balance at every level of the layout. The rationale behind this is that we have many operators at each level, and each of them is not expensive to execute. Given that, we can solve the relaxed problem and round the solution to integer values. The total network time of each container at level $i$ is computed as:
$$\overrightarrow{NET_H}[i] / (net\_speed \cdot min(\overrightarrow{L_P}[i-1], \overrightarrow{L_P}[i])$$
since the maximum network throughput between two consecutive level $i$-$1$ and $i$ is determined by the minimum number of containers in these two levels.

We separate the prediction window into two parts: the first involving re-organization along with query execution (denoted as $t_P^d$) and the second involving query execution only. The length of the first period is estimated by the time needed to perform data re-organization using the model of $size_d(x,y)$ defined above as follows:
$$t_P^d = \frac{size_d(\overrightarrow{L_H}[1], \overrightarrow{L_P}[1])}{|\overrightarrow{L_H}[1] -  \overrightarrow{L_P}[1]| \cdot Arc \cdot net\_speed}$$
where ($|\overrightarrow{L_H}[1] -  \overrightarrow{L_P}[1]| \cdot Arc$) is the number of containers \emph{Arc} in the circle affected by the change. These containers will transfer table partitions from the cloud storage through the network with $net\_speed$ being the network speed. Thus, the length of the second period exclusively dedicated to query processing is $W_P - t_P^d$. Notice that the faster the time to re-organize the data is, the longer the period of time spent to execute queries becomes. This is the reason why our method prefers to perform changes to the number of data containers that are near the diagonal as shown in Figure~\ref{fig:data-move-model}(a).

Our modeling could potentially include in the re-organization part the time to create a \vm\ and initialize it. A simple approach would be to consider this time a constant (e.g., 1 minute). However, in most clouds, this is relatively small compared to the actual time that the \vms\ are used~\cite{Amazon:WS} and some cloud providers allow for pre-configured instances\footnote{Okeanos: okeanos.grnet.gr, \\ eCloudManager: www.fluidops.com} which can be created in seconds, making the initialization time negligible. Most importantly however, changing the shape of the virtual infrastructure does not directly imply the allocation of new \vms . In our implementation, containers scheduled to be deleted, are kept until their entire quantum has finished. If the virtual infrastructure needs to grow in size, we opportunistically re-use any available containers from those scheduled to be deleted, and essentially eliminate their initialization cost.

We compute the estimated number of queries per \sla\ in each of the two parts of the prediction window as follows:
$$\overrightarrow{Q_P^d} = \overrightarrow{Q_H} \cdot t_P^d / W_H$$
$$\overrightarrow{Q_P} = \overrightarrow{Q_H} \cdot (W_P -  t_P^d) / W_H$$
Using the estimated number of queries, the predicted revenue per \sla\ class for the two part of the prediction period is as follows:
$$\overrightarrow{R_P^d} = \overrightarrow{Q_P^d} \cdot \overrightarrow{\alpha} \cdot e^{(-(t_P^d + t_P) / \overrightarrow{\gamma})}$$
$$\overrightarrow{R_P} = \overrightarrow{Q_P} \cdot \overrightarrow{\alpha} \cdot e^{(-t_P / \overrightarrow{\gamma})}$$
Notice that we include the time to perform data re-organization $t_P^d$ in the calculation of the revenue in the first period ($\overrightarrow{R_P^d}$) of the prediction window. The total revenue in the prediction window is as follows:
$$R = \sum_i(\overrightarrow{R_P^d}[i]) + \sum_i(\overrightarrow{R_P}[i])$$

The operational cost is computed by adding the time quanta $T_Q$ of the allocated containers in the prediction window $W_P$ and multiplying by the quantum cost $M^c_Q$ as:
$$O = M^c_Q \cdot \frac{W_P}{T_Q} \sum_i(\overrightarrow{L_P}[i])$$

The profit generated is computed as $R-O$. We seek to find $\overrightarrow{L_P}$ that maximizes profit.Since the number of container layouts is limited assuming a maximum number of containers per level (e.g., $100$), we could potentially compute the revenue enumerating all different layouts. The total number of layouts with height $4$ and a maximum of $100$ containers/level is $10^{8}$. In practice, this number is infeasible to compute exhaustively. Instead, we maximize the profit function using the {\sl L-BFGS-B} Algorithm~\cite{Byrd:1995} which is a general purpose iterative optimization method that finds local maxima/minima of multivariable functions. Since the {\sl L-BFGS-B} finds solutions with real numbers, we round the solutions to the ceiling (e.g., a value of $13.4$ becomes $14$ containers).

We seed {\sl L-BFGS-B} with the previous layout ($\overrightarrow{L_H}$) as the starting point. Extensive experimentation through enumeration of all solutions and comparison of outcomes to those derived with the help of {\sl L-BFGS-B} showed that solutions are very close (yet, they are not identical due mostly to rounding). This was expected as changes is the topology are mostly gradual because of the data re-organization cost. The seeding the {\sl L-BFGS-B} with the previous container layout ($\overrightarrow{L_H}$) is sufficient to adequately guide the algorithm.

\subsection{Query Tree Scheduler}
The execution tree plan is scheduled by performing load balance on every level of the layout while considering current load at each container. The load is quantified as the number of running and queued operators. First, we find the {\bf rank} of each operator that is the height of the node in the execution tree (Figure~\ref{fig:tree-exec-plan}). The rank of an operator determines the level of the layout at which is scheduled. As there is at least one container allocated in each level, we can always find at least one valid schedule. Once we determine the levels in which all operators are placed, we order containers at each level according to their load. The scheduler maps the operators of the each level of the query tree to the corresponding containers using the increasing ordering in a round robin fashion. For generic dataflow graphs, the scheduling problem is a much harder and more advanced methods should be used~\cite{DBLP:conf/sigmod/KllapiSTI11}. However, in this work we consider only tree--query plans. The specialized scheduling algorithm discussed here works because of the following two reasons: 
\begin{inparaenum}[\itshape i\upshape)]
\item individual operators are not expensive to execute and 
they do not generate voluminous data as they use aggregate functions. 
This has as a consequence that even sub-optimal assignments of operators 
will not cause much imbalance, 
\item
operators that are at the same level of the execution tree, 
will have approximately the same execution time since the data 
is balanced.
\end{inparaenum}

Our scheduling method is robust to use in practice since it neither assumes a particular operator behavior nor uses a model to predict execution times. The elastic layout allocation algorithm exclusively uses historical measurements taken after queries have been executed and so actual running times of their operators are known. Further, ongoing queries are not affected by changes in the container layout as partitions located at the respective \vms\ are not deleted even if they are re-assigned elsewhere. This is possible because of the de-coupled nature of the used compute and storage resources. Finally, our proposed algorithm is ideal when used for queries featuring \udfs\ unknown properties. \udfs\ are encountered frequently and their modeling and behavior prediction remains an open problem.

%% file: experiments.tex
%!TEX root = ElasticTreePaper.tex

\section{Experimental Evaluation}
\label{sec:exps}

The objectives of our experimentation are to:
 \begin{inparaenum}[\itshape A\upshape)]
\item evaluate our engine and show that we can achieve near-interactive response times for analytical queries,
\item show that we can efficiently execute complex analytical queries with \udfs\ that have arbitrary user code, and 
\item examine the effectiveness of the proposed elastic container layout algorithm and ascertain its ability to adapt to the workload. 
\end{inparaenum}

\subsection{Experimental Setup}
\label{sec:exp-setup}

\noindent\textbf{Experimental Environment:}
We have implemented the functionality presented within {\sysname}~\cite{DBLP:journals/debu/TsangarisKKPPPSSI09}, our system for dataflow execution on the cloud. We compare our approach with the latest version of \emph{Cloudera Impala}, the state-of-the-art in-memory analytics platform~\cite{impala}. We deployed the systems in the {\sl Okeanos} cloud~\footnote{\url{okeanos.grnet.gr}} and used up to $64$ \vms\  for processing, each with $1$ CPU, $4$ GB of memory, and $20$ GB of disk. We measured the network bandwidth to be around $150$ Mbps. We set the quantum $T_Q$ to $300$ seconds and the cost of the quantum $M_Q^c$ to \$0.41 (or equivalently $\sim$\$5/hour). The memory of the operators in the execution tree is set to 10\% of the container's memory, i.e., at most 10 {\em leaf}, {\em internal}, or {\em root} queries can run concurently in each container. We also used a latest version of the HDFS distributed file system\footnote{HDFS version 2.6 \url{hadoop.apache.org}} as a storage service deployed in 8 \vms\ to store table partitions.

\noindent\textbf{Datasets:} we used two datasets namely, {\sl TPC-H}~\cite{tpch} that typically models data warehouse settings, and {\em Freebase}, an {\sl RDF} dataset\footnote{\url{developers.google.com/freebase/data}}. The {\sl TPC-H} benchmark has eight tables:

\noindent
{\small
$lineitem(128, l\_orderkey),
orders(128, o\_orderkey),
part(1), \\
partsupp(1),
supplier(1),
customer(1),
region(1),
nation(1)$
}

\begin{figure} [t!]
\centering
\includegraphics[width=8.5cm]{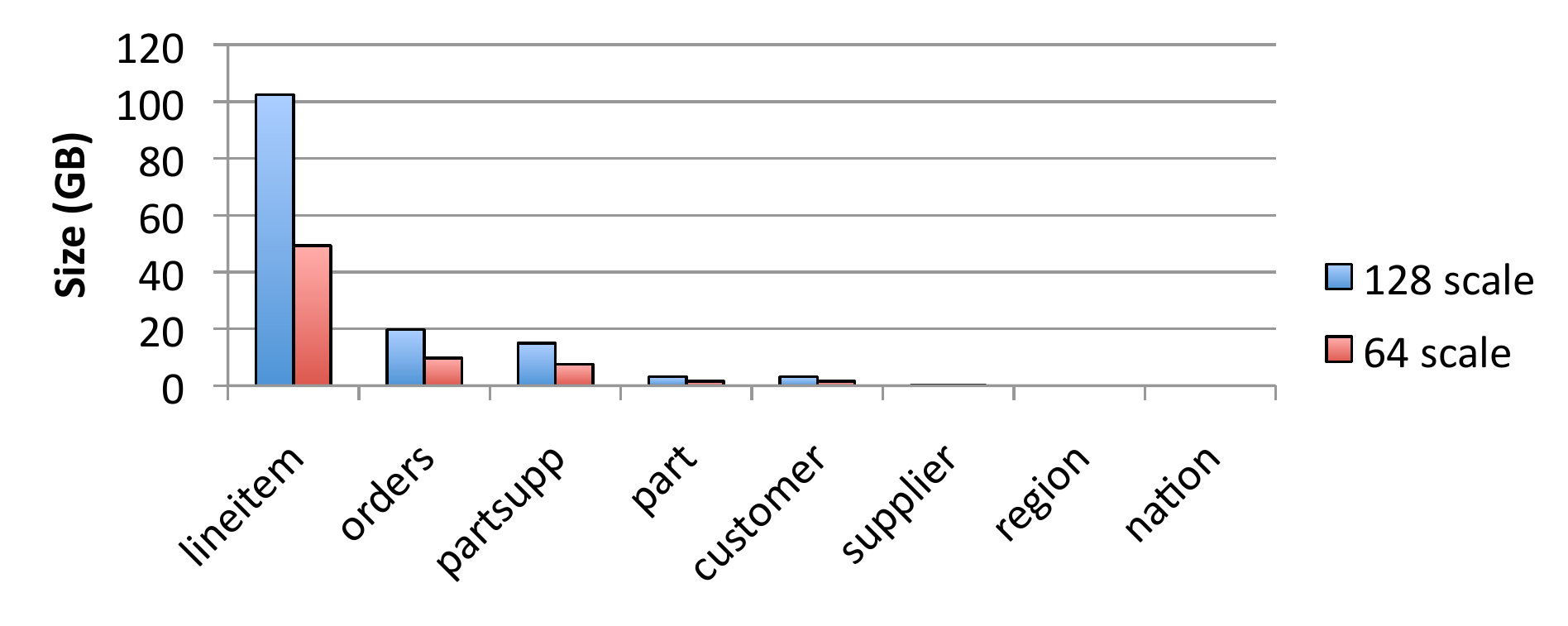}
\vspace{-0.5cm}
\caption{{\sl TPC-H} table size distribution at $64$ GB and $128$ GB scales.}
\label{fig:tpchdist}
\end{figure}

In parentheses, we indicate the number of partitions we have created for each table and the key(s) based on which we performed table partitioning. We partition tables \emph{lineitem} and \emph{orders} on their foreign key using hash partitioning and replicate all other (smaller) tables. We used the 8 ($\sim$8GB),  64 ($\sim$64GB), and 128 ($\sim$128 GB) as the {\sl TPC-H} scale--factors. Figure~\ref{fig:tpchdist} shows the sizes of the benchmark tables illustrating the large size difference between the fact table \emph{lineitem} and the rest of the tables.

{\em Freebase} contains approximately $2.5$ billion tuples in the form of {\sl RDF} triples: 
{\em $<$subject$>$ $<$predicate$>$ $<$object$>$ ``.''} and its volume stands at $250$ GB. If the object is text, it is tagged at its end with the appropriate language symbol (e.g., {\sf @en} means text in English). We load {\em Freebase} data into a $3$-column table.

\noindent\textbf{Queries:} we use a subset of the {\sl TPC-H} queries that cover a wide range of the types of queries we target. In particular, we choose queries 1, 3, 4, 5, 7, and 9. 1 uses only table $lineitem$ and has 8 aggregate functions. Queries 3 and 4 have a small number of joins (less than $3$) and a small number of aggregate functions while queries 5, 7, and 9 feature a large number of joins and several aggregate functions. With {\em Freebase}, we utilize two queries with complex \udfs\ to create a histogram of the languages that appear in the dataset. The {\em first} query uses regular expressions to separate the language of each object and then counts the number of languages encountered. The query is as follows:

\begin{small}
\begin{verbatim}
SELECT lang, count(lang) as c
FROM (SELECT REGEXPR('.*@(.*)', o) as lang
      FROM freebase WHERE o like "%@%")
GROUP BY lang ORDER BY c desc;
\end{verbatim}
\end{small}

The {\em second} query uses \emph{reservoir sampling} to sample $1$ million rows from the table and computes the histogram though a \udf\ that is applied on the sample and detects the language of a given text using a statistical model. The query is the following:

\begin{small}
\begin{verbatim}
SELECT lang, count(lang) as c
FROM (SELECT DETECTLANG(sobj) as lang
      FROM (SELECT SAMPLE(1000000, obj) as sobj
            FROM freebase))
GROUP BY lang ORDER BY c desc;
\end{verbatim}
\end{small}

\noindent{\bf SLAs and Query Generator Client:} We use two types of SLAs: ``normal'' with $\alpha=10$ \& $\gamma=80$ and ``high priority'' with $\alpha=20$ \& $\gamma=40$. We also created a generator that launches queries with a Poisson distribution. More specifically, the generator computes the arrival time $k$ (in seconds) of the next query as $f(k; \lambda)= \Pr(X=k)= \lambda^k e^{-\lambda} / k!$, where $\lambda$ is the expected value of $X$ (in seconds). We can achieve desired query rates by setting $\lambda$ appropriately. For example, if $\lambda=10$, one query is issued to the engine every $10$ seconds on average.

\noindent{\bf Algorithms and Measurements:} 
We use our elastic \vm\ layout allocation algorithm to adjust the size of the virtual infrastructure. As a baseline, we  select a {\em static layout} that remains fixed over time. We use two such static allocations:
{\em small} with ($10$, $4$, $1$) and
{\em large} ($42$, $12$, $3$); here, we designate within parentheses the number of containers per layout level starting from the lower level $L_0$ that contains the data. We bootstrap our dynamic layout allocation algorithm with a {\em medium} static configuration ($26$, $8$, $2$). Finally while experimenting, we measure the following: average execution time for queries, revenue, cost, and average number of \vms\ used at each layout level.

\subsection{Near-Interactive Analytics}
\begin{figure} [t!]
\centering
\includegraphics[width=8.5cm]{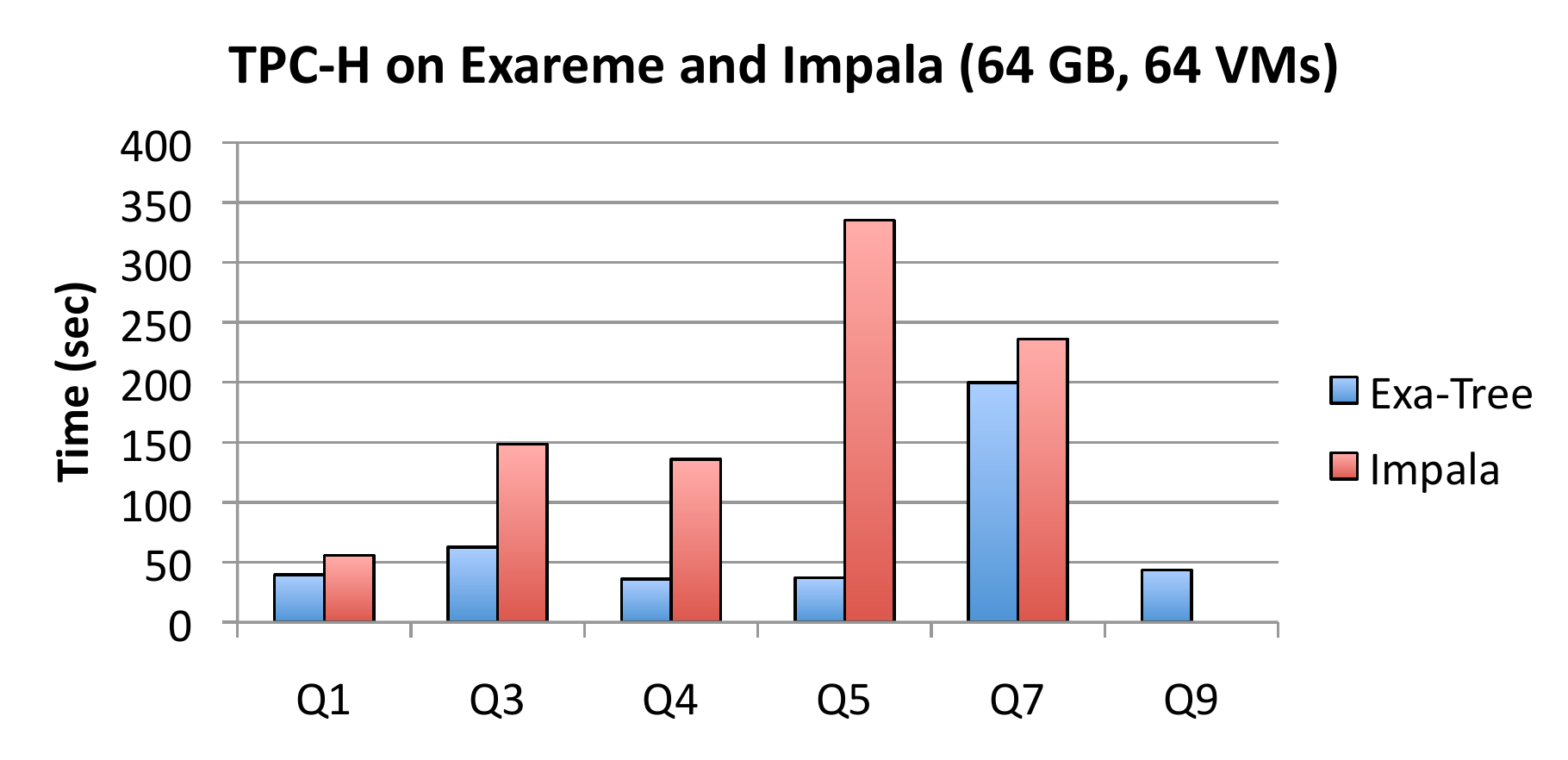}
\vspace{-0.5cm}
\caption{{\sl TPC-H} with $64$ GB on \emph{Impala} and \emph{Exareme} using $64$ \vms.}
\vspace{-0.3cm}
\label{fig:systems_compare}
\end{figure}

In our first set of experiments, we validate the efficiency of the system by executing a single type of query at a time and measuring corresponding turnaround time. We run each query $4$ times and report the average of the last $3$ measurements, a technique  also followed by others~\cite{DBLP:conf/sigmod/XinRZFSS13}. In this way, the observed execution times reflects the behavior of the system in live operation. We use the {\sl TPC-H} benchmark with $64$ \vms\  on {\sl Okeanos} and a $3$-level execution tree. Figure~\ref{fig:systems_compare} compares performance of our implementation, termed \emph{Exa-Tree}, with that of \emph{Impala} while using $64$ GB of data on $64$ \vms. We observe that  \emph{Exa-Tree} is comparable, and in some cases more efficient, for the types of queries we focus on in this work. This is due to our data partitioning and placement scheme that reduces network traffic during query execution (due to replication) and the tree execution plans. As \emph{Impala} runs entirely in memory, we were not able to run query 9 because we reached memory limits.

\begin{figure} [h!]
\centering
\includegraphics[width=8.5cm]{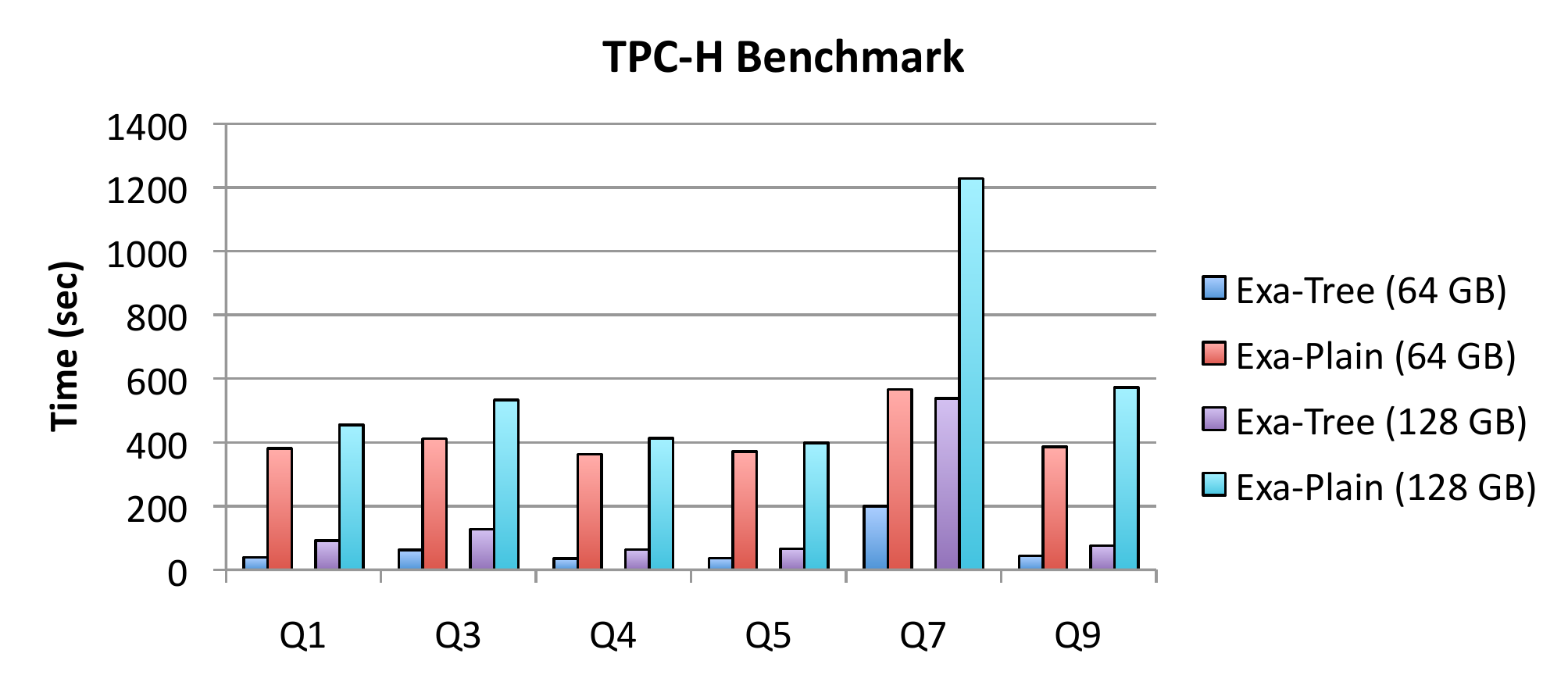}
\vspace{-0.5cm}
\caption{TPC-H queries using tree and graph execution plans on 64 containers.}
\label{fig:efficiency}
\end{figure}

We also compared with a previous version of Exareme that used graphs to execute queries. Figure~\ref{fig:efficiency} shows the results. We observe that queries executed using tree execution plans run significantly faster. The main reason is the tree execution in combination with the exploitation of data partitioning. The previous version of the system used a lattice (all-to-all connections) to partition the data and perform aggregations in parallel. Using tree execution plans, we radically reduce the number of connections, improving the system performance by up to an order of magnitude and offer near-interactive response times (as small as 35 seconds on the 64 GB scale).

\subsection{Complex Analytics}
\begin{figure} [t!]
\centering
\includegraphics[width=7.0cm]{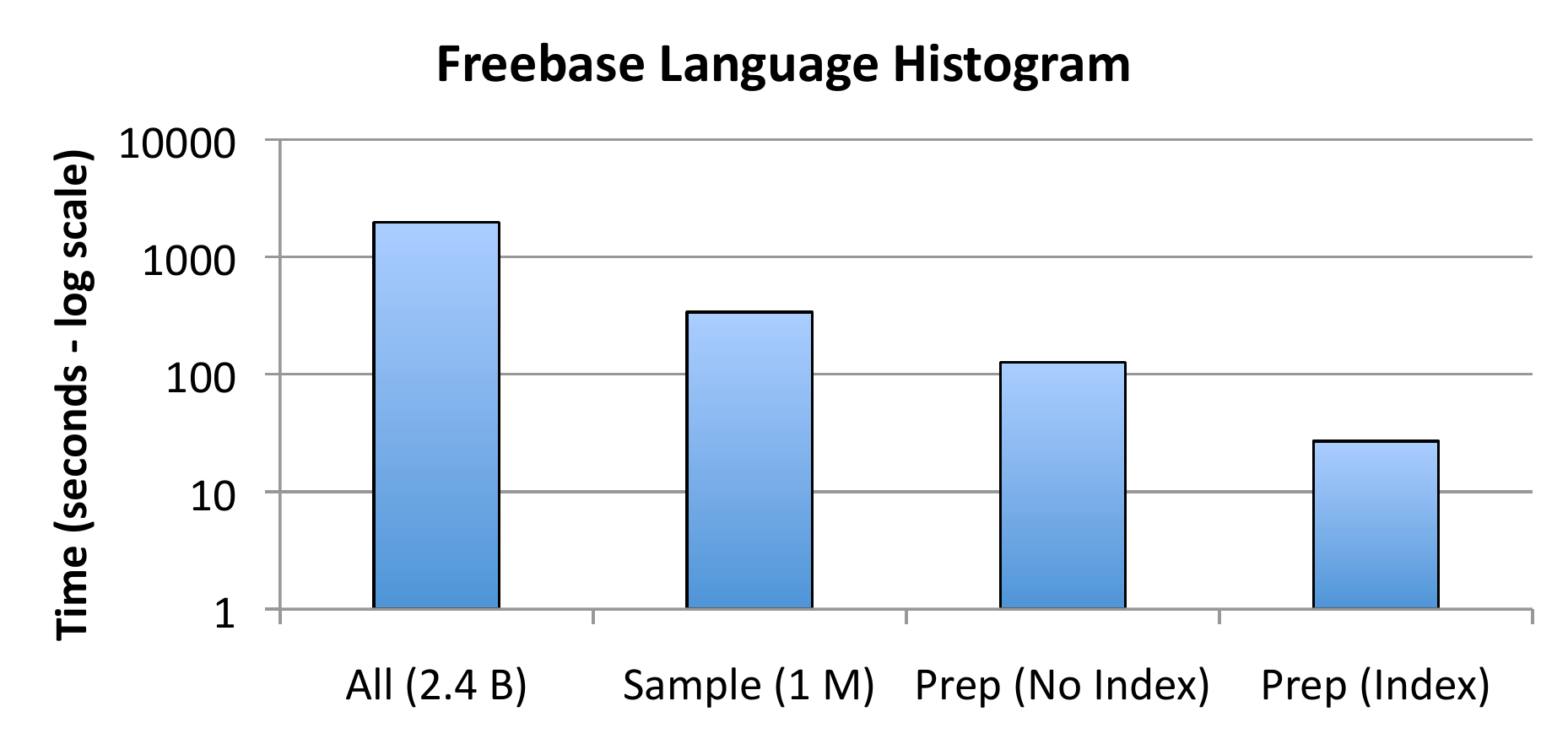}
\vspace{-0.4cm}
\caption{Execution times for \emph{Freebase} queries.}
\label{fig:freebase}
\end{figure}

In the second set of experiments, we assess the efficiency of our engine on complex analytics expressed in \udfs, again by executing a single query at a time and measuring respective execution times. As previously, we run each query 4 times and report the average of the last 3 times.  We use the {\em Freebase} dataset and the two queries mentioned earlier in the section using $64$ \vms. Figure~\ref{fig:freebase} depicts the attained execution times for the two queries ({\sl All} and {\sl Sample}). In the first query ({\sl All}), operators at the leaves of the execution tree take most of the time as computing 2.4 billion regular expressions is expensive. The second query ({\sl Sample}) being highly selective completes in $339$ seconds. It is worth mentioning that both queries produce  similar distributions as shown in Table~\ref{tab:freebase_result}. We also pre-processed the $<$\emph{object}$>$-column by extracting the language tag and created an additional column on the table hosting the \emph{Freebase}. Here, the histogram on the entire dataset is computed in merely $107$ seconds without indexes and in $27$ seconds using indexes. This performance highlights the near--real-time capabilities of our engine in large datasets.

\begin{table}[h!]
\centering
    \begin{small}
    \begin{tabular}{|l|r|l|r|}
    \hline
 \multicolumn{2}{|c|} {\bfseries All (2.4B)} & \multicolumn{2}{c|} {\bfseries Sample (1M)}  \\
    \hline
    \hline
{\bf lang} & {\bf count} & {\bf lang} & {\bf count} \\
    \hline
    \hline
en & 134096634 & en & 115335 \\
    \hline
fr	 & 28091737 & fr  & 23991 \\
    \hline
de &  27890842 & de & 23906 \\
    \hline
es	 & 26934217 & es  & 23462 \\
    \hline
it	&  26516667 & it  & 23148 \\
    \hline
... &  ... & ...  & ... \\
    \hline
    \end{tabular}
    \end{small}
\caption{Freebase Language Histogram}
\label{tab:freebase_result}
\end{table}

% This is from layout stabilization
\begin{figure*} [t!]
\centering
\includegraphics[width=18cm]{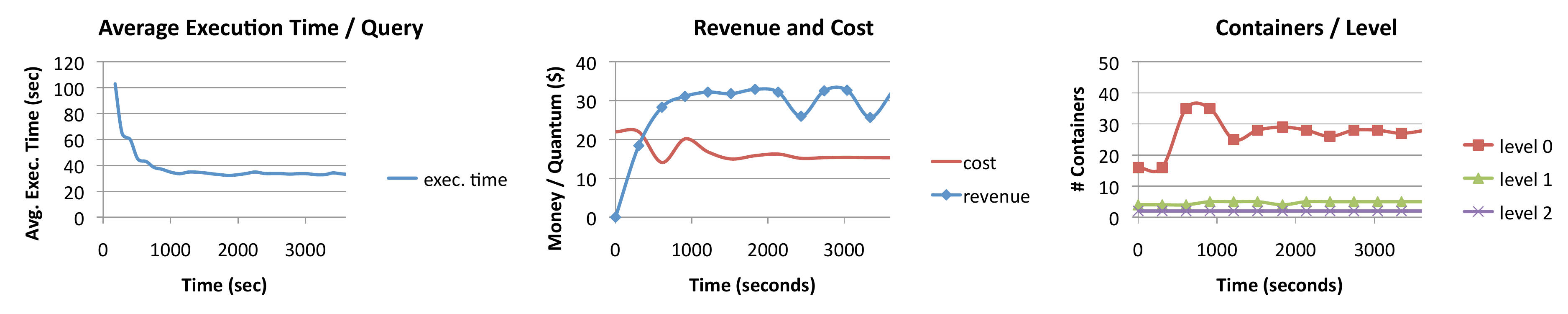}
\vspace{-0.5cm}
\caption{Query exec. time (left), revenue \& cost (middle), and containers allocated per level (right).}
\label{fig:stable-exectime}
\end{figure*}

\subsection{Elasticity under Dynamic Workloads}
In this set of experiments, we examine both the effect that the elasticity has on query execution time and the profit generated. For these experiments we used TPC-H with scale factor 8 in order to be able to run the queries with a variety of infrastructure sizes. The clients connected to the system issue the queries 1 and 3 of the benchmark.

\subsubsection{Layout Stabilization}
Here, we examine the stabilization of the virtual infrastructure. We use a workload with Q1 using the ``normal'' \sla\  and Poisson parameter $\lambda = 60$.  The left part of Figure~\ref{fig:stable-exectime} shows the average execution time of the queries over time. We observe that our algorithm is able to stabilize quickly after 800 seconds. The delay to reach steady state in the beginning is due to the initial data transfer from the cloud storage to the containers. This explains the high average running time of the queries during that period. The middle part of Figure~\ref{fig:stable-exectime}, shows the revenue and the corresponding cost of the allocated virtual infrastructure. In the beginning, the revenue is actually lower than the cost, and thus, there is a loss instead of profit. After the data transfer has finished, the profit is stabilized at a significantly high value. Finally, the right part of Figure~\ref{fig:stable-exectime} shows the number of containers at each level of the layout over time. We observe that the virtual infrastructure adapts to the workload taking the shape of a tree, with most of the allocated containers located at the data level.

\subsubsection{Compare with Static Infrastructures}
\begin{figure} [t!]
\centering
\includegraphics[width=8.0cm]{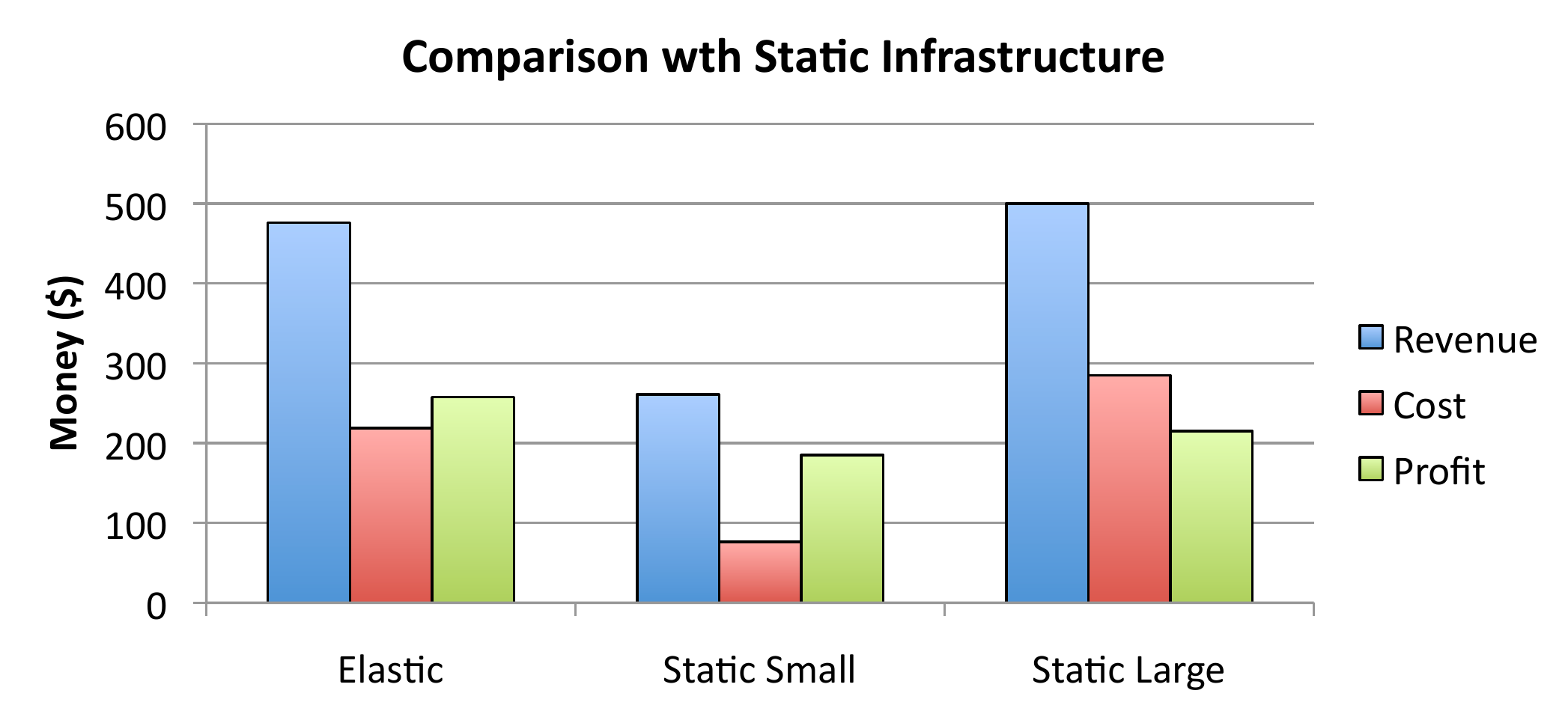}
\vspace{-0.2cm}
\caption{Elastic configuration vs. static layouts.}
\label{fig:elastic-compare}
\end{figure}

Figure~\ref{fig:elastic-compare} depicts the profit gained when the static \vm\ configurations are used to handle the workload as well as the profit generated by our approach. We run the system for one hour using a client that issues query $Q1$ in three phases, each of 20 minute duration. In the first and third phase, the Poisson parameter $\lambda$ is set to 60 and in the second phase to 30 (the rate is doubled).

We readily ascertain that smaller-sized infrastructures produce less revenue as expected. Similarly, the expended costs increase as more \vms\ and time quanta are used. The elastic layout allocator however produces a better-fitted layout that adapts to the workload changes and yields the highest profit compared to all static choices. Lastly, the elastic approach does generate less revenue than the \emph{large} infrastructure. However, this is in sequence with our design as we optimize for profit and not for revenue.

\subsubsection{Measure adaptivity with Dynamic Workload}
In our final set of experiments, we evaluate the adaptability of our elastic online algorithm in presence of workloads whose features change over time. In particular, we employ a workload consisting of three stages, of 1 hour each, where query workload characteristics are perturbed between the stages. As a default workload we issue Q1 with a Poisson parameter $\lambda=60$ and using the ``normal'' \sla\ . We change this default query workload in the second stage using the following three options:

\begin{figure*} [t]
\centering
\includegraphics[width=18.0cm]{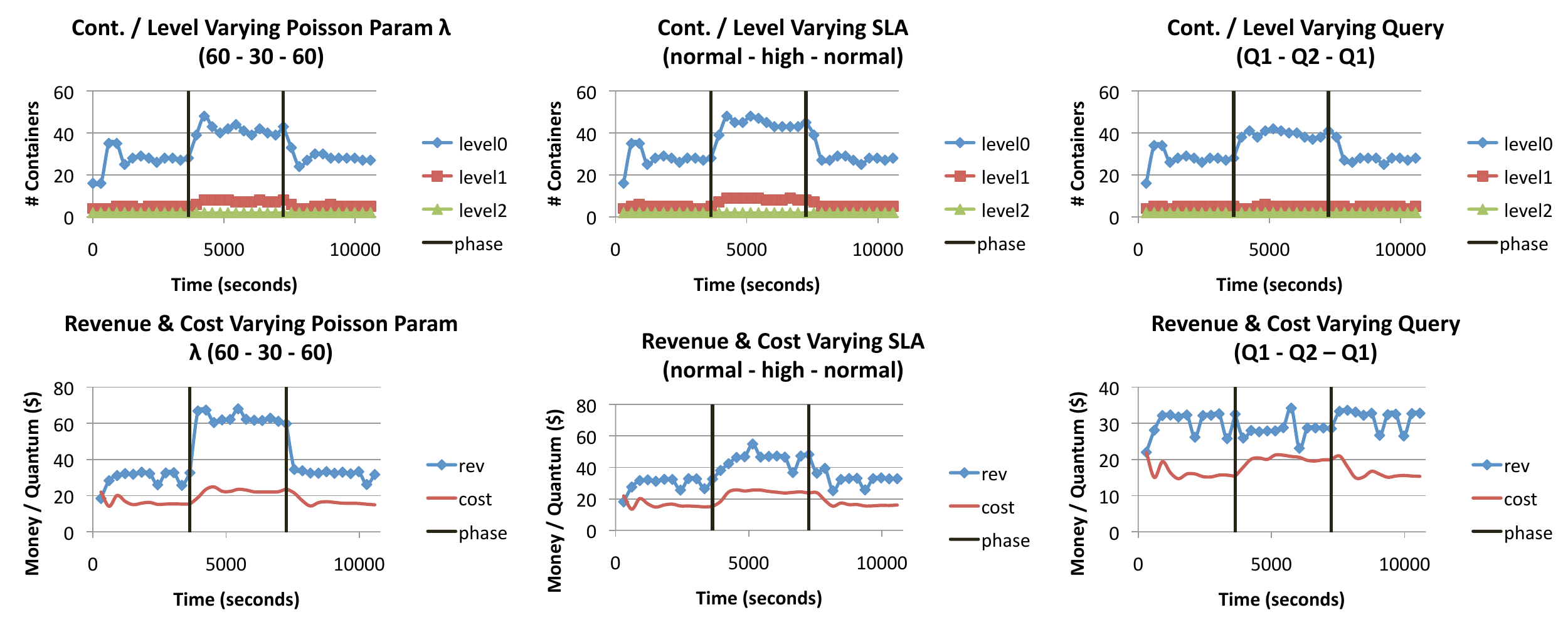}
\vspace{-0.7cm}
\caption{Elastic containers allocated per tree level and revenue and cost for workload with different phases.}
\label{fig:phase-workload}
\end{figure*}

\noindent{$\bullet$ \bf Varying Query Rates:} we vary the rate with which queries are issued by setting the Poisson parameter to $\lambda=30$ in the second stage and essentially, doubling the rate. The left part of Figure~\ref{fig:phase-workload} shows the \vms\  allocated per layout level as well as revenue. Our approach does rapidly adapt to varying workload and starts adjusting the number of \vms\ exactly at the phase boundaries. We also observe the number of containers allocated is increased along with the query rate as more revenue is generated.

\noindent{$\bullet$ \bf Varying SLAs:} we vary the \sla\ type to ``high priority'' during stage 2, while phases 1 and 3 have queries with the ``normal'' \sla. The middle part of Figure~\ref{fig:phase-workload} shows our execution results: for queries with a higher price, our algorithm designates more \vms\ to generate additional revenue.  

\noindent{$\bullet$ \bf Varying Query:} in our final experiment, we vary the type of the queries issued. In stages 1 and 3, we use $Q1$ and in stage 2 we use $Q3$. The right part of Figure~\ref{fig:phase-workload} shows once again the superiority of the elastic algorithm when it comes to the rapid adaption of the virtual infrastructure. We observe that for $Q3$ the profit drops because it is more expensive to execute. The algorithm allocates more containers in order to be able to keep the profit positive.

%% file: related.tex
%!TEX root = ElasticTreePaper.tex

\section{Related Work}
\label{sec:related}

There are several areas of data management where related work has been conducted. We briefly outline here key results from the fields of data warehouses, {\sl NoSQL}-systems, and elasticity.

\subsection{Data Warehouses}
Data Warehouses store very large volumes of data and are typically used for report generation and historical analyses to discover trends. Several systems have been implemented that are open--source (e.g., \emph{Hive}~\cite{DBLP:conf/icde/ThusooSJSCZALM10}), proprietary (e.g., \emph{Tenzing}~\cite{DBLP:journals/pvldb/ChattopadhyayLLMALKW11}), or commercial (e.g.,\emph{Vertica}~\cite{DBLP:journals/pvldb/LambFVTVDB12}). The most popular open--source warehouses are based on \emph{MapReduce}~\cite{DBLP:conf/osdi/DeanG04, DBLP:conf/icde/ThusooSJSCZALM10} and typically offer high level languages (e.g., {\sl SQL}) to express queries. The latter are ultimately transformed to one or more \emph{MapReduce} jobs~\cite{DBLP:conf/icdcs/LeeLHWHZ11}.  The \emph{MapReduce} abstraction however is not efficient for heavy aggregate queries that we target in this work. In \emph{MapReduce}, multi-level aggregations can only be expressed using multiple jobs, rendering the approach less efficient than that of a tree abstraction. Moreover, the optimization goal of these systems is to both minimize the number of jobs they produce as well as to maximize parallelization in order to minimize their total execution time. The monetary cost of the resources is by and large ignored. The same holds for \emph{Dremel}~\cite{DBLP:journals/pvldb/MelnikGLRSTV10} and \emph{Scuba}~\cite{DBLP:journals/pvldb/AbrahamABBCGMMRSWZ13} which has been recently proposed as specialized systems targeting query--tree executions, and, furthermore, to the best of our knowledge, are not elastic.

\subsection{{NoSQL}--Systems}
Several systems have been proposed to manage data in formats different than relational tables. Examples include \emph{MongoDB}~\cite{DBLP:books/daglib/0025185}, \emph{Sawzall}~\cite{DBLP:journals/sp/PikeDGQ05}, \emph{PigLatin}~\cite{DBLP:conf/sigmod/OlstonRSKT08}, and \emph{FlumeJava}~\cite{Chambers:2010}. All of the above are built either on top of \emph{MapReduce} and so, they inherit all pertinent weaknesses mentioned earlier, or built from scratch by following approaches that are not suitable for the queries we target here~\cite{DBLP:books/daglib/0025185}. Furthermore, no such system offers a clean and simple way to define new UDFs and their properties so that they may be used during optimization.

\subsection{Elasticity}
Several works focus on cloud elasticity~\cite{DBLP:conf/edbt/TinnefeldKGBRSP13, DBLP:conf/icde/TsakalozosKSRPD11, trfd-icdcs10}, and dynamically allocating resources to increase performance. A recent work~\cite{DBLP:conf/sigmod/SchaffnerJKKPFJ13} focuses on how to minimize the number of \vms\ used to save on cost, but this is not a plausible strategy in our setting where queries are associated with \slas\ and the goal is to maximize profit. Some works examine cloud elasticity in the context of in-memory distributed transactions~\cite{DBLP:journals/tods/DasAA13}. In our setting, the data are updated using bulk loading every day or week.

Elasticity for array databases is examined recently~\cite{DBLP:conf/sigmod/DugganS14}. This work, similarly to our methodology, makes predictions about the future based on past queries. However, the proposed algorithm is only applicable to array-based scientific data (that only grow in size and rarely deleted) and considers only increasing the size of the virtual infrastructure. We focus on a more generic problem.

To the best of our knowledge, none of the proposed solutions is suitable for our setting. Our proposal exploits cloud elasticity by automatically adjusting the size of the allocated virtual infrastructure to maximize profit by taking into account \slas\ and the monetary cost for using cloud resources that has been in general ignored thus far.

%% file: conclusions.tex
%!TEX root = ElasticTreePaper.tex

\section{Conclusions}
\label{sec:future}

We propose an elastic engine built on top of \iaas\ clouds to execute queries with a tree execution plan encountered in a large set of analytical {\tt SQL} queries that involve heavy aggregations. We suggest to layout the allocated infrastructure  \iaas\ nodes in a tree shape so that we can naturally map the execution plans of these queries. Our elastic \vm\ allocation algorithm dynamically changes the container layout based on the query workload monitored over a sliding time window. Our objective is to maximize the profit generated taking into account the monetary cost of the resources as well as the revenue generated by the query workload. Finally, we shown that our approach offers near-interactive response times and adapts quickly to workload changes.

%% file: ElasticTreePaper.bbl
\begin{thebibliography}{10}

\bibitem{impala}
"{Cloudera Impala}: Open source, interactive {SQL} for {Hadoop},
  http://www.cloudera.com/content/cloudera/en/
  products-and-services/cdh/impala.html".

\bibitem{tpch}
{TPC-H} {B}enchmark, \url{http://www.tpc.org/tpch/}.

\bibitem{DBLP:journals/pvldb/AbrahamABBCGMMRSWZ13}
L.~Abraham, J.~Allen, O.~Barykin, V.~R. Borkar, B.~Chopra, C.~Gerea, D.~Merl,
  J.~Metzler, D.~Reiss, S.~Subramanian, J.~L. Wiener, and O.~Zed.
\newblock Scuba: Diving into data at facebook.
\newblock {\em {PVLDB}}, 6(11):1057--1067, 2013.

\bibitem{Amazon:WS}
Amazon.
\newblock {Web Services}, http://aws.amazon.com.

\bibitem{DBLP:journals/tcst/AngCL05}
K.~H. Ang, G.~Chong, and Y.~Li.
\newblock {PID} control system analysis, design, and technology.
\newblock {\em {IEEE} Trans. Contr. Sys. Techn.}, 13(4):559--576, 2005.

\bibitem{apache:hadoop}
Apache.
\newblock Apache hadoop, http://hadoop.apache.org/.

\bibitem{DBLP:journals/cacm/ArmbrustFGJKKLPRSZ10}
M.~Armbrust, A.~Fox, R.~Griffith, A.~D. Joseph, R.~H. Katz, A.~Konwinski,
  G.~Lee, D.~A. Patterson, A.~Rabkin, I.~Stoica, and M.~Zaharia.
\newblock A view of cloud computing.
\newblock {\em Commun. ACM}, 53(4):50--58, 2010.

\bibitem{Byrd:1995}
R.~H. Byrd, P.~Lu, J.~Nocedal, and C.~Zhu.
\newblock A limited memory algorithm for bound constrained optimization.
\newblock {\em SIAM J. Sci. Comput.}, 16(5):1190--1208, Sept. 1995.

\bibitem{Chambers:2010}
C.~Chambers, A.~Raniwala, F.~Perry, S.~Adams, R.~R. Henry, R.~Bradshaw, and
  N.~Weizenbaum.
\newblock Flumejava: easy, efficient data-parallel pipelines.
\newblock In {\em {PLDI} 2010, Toronto, Ontario, Canada, June 5-10, 2010},
  pages 363--375, 2010.

\bibitem{DBLP:journals/pvldb/ChattopadhyayLLMALKW11}
B.~Chattopadhyay, L.~Lin, W.~Liu, S.~Mittal, P.~Aragonda, V.~Lychagina,
  Y.~Kwon, and M.~Wong.
\newblock Tenzing {A} {SQL} implementation on the mapreduce framework.
\newblock {\em {PVLDB}}, 4(12):1318--1327, 2011.

\bibitem{DBLP:books/daglib/0025185}
K.~Chodorow and M.~Dirolf.
\newblock {\em MongoDB - The Definitive Guide: Powerful and Scalable Data
  Storage}.
\newblock O'Reilly, 2010.

\bibitem{DBLP:journals/tods/DasAA13}
S.~Das, D.~Agrawal, and A.~{El Abbadi}.
\newblock Elastras: An elastic, scalable, and self-managing transactional
  database for the cloud.
\newblock {\em {ACM} Trans. Database Syst.}, 38(1):5, 2013.

\bibitem{DBLP:conf/osdi/DeanG04}
J.~Dean and S.~Ghemawat.
\newblock {MapReduce: Simplified Data Processing on Large Clusters}.
\newblock In {\em OSDI},
  pages 137--150, 2004.

\bibitem{DBLP:conf/sigmod/DugganS14}
J.~Duggan and M.~Stonebraker.
\newblock Incremental elasticity for array databases.
\newblock In {\em {SIGMOD} 2014, Snowbird, UT, USA, June, 2014}, pages
  409--420, 2014.

\bibitem{DBLP:journals/sigmod/FlorescuK09}
D.~Florescu and D.~Kossmann.
\newblock Rethinking cost and performance of database systems.
\newblock {\em SIGMOD Record}, 38(1):43--48, 2009.

\bibitem{DBLP:conf/sosp/GhemawatGL03}
S.~Ghemawat, H.~Gobioff, and S.-T. Leung.
\newblock The google file system.
\newblock In {\em SOSP}, pages 29--43, 2003.

\bibitem{book-big-data-2013}
M.~Giese, D.~Calvanese, P.~Haase, I.~Horrocks, Y.~Ioannidis, H.~Kllapi,
  M.~Koubarakis, M.~Lenzerini, R.~M{\"o}ller, M.~Rodriguez-Muro,
  {\"O}.~{\"O}zcep, R.~Rosati, R.~Schlatte, M.~Schmidt, A.~Soylu, and
  A.~Waaler.
\newblock Scalable end-user access to big data.
\newblock In R.~Akerkar, editor, {\em Big Data Computing}. CRC Press, 2013.

\bibitem{DBLP:journals/ccr/GonzalezMCL09}
L.~M.~V. Gonzalez, L.~R. Merino, J.~Caceres, and M.~Lindner.
\newblock {A break in the clouds: towards a cloud definition}.
\newblock {\em Computer Communication Review}, pages 50--55, 2009.

\bibitem{DBLP:books/mk/HanK2000}
J.~Han and M.~Kamber.
\newblock {\em Data Mining: Concepts and Techniques}.
\newblock Morgan Kaufmann, 2000.

\bibitem{DBLP:conf/stoc/KargerLLPLL97}
D.~R. Karger, E.~Lehman, F.~T. Leighton, R.~Panigrahy, M.~S. Levine, and
  D.~Lewin.
\newblock Consistent hashing and random trees: Distributed caching protocols
  for relieving hot spots on the world wide web.
\newblock In {\em STOC, El Paso, Texas, USA, May 4-6}, pages 654--663, 1997.

\bibitem{DBLP:conf/owled/KllapiBHIJKKZ13}
H.~Kllapi et~al.
\newblock Distributed query processing on the cloud: the optique point of view
  (short paper).
\newblock In {\em {OWLED}, Montpellier, France, 2013.}, 2013.

\bibitem{saj}
H.~Kllapi, B.~Harb, and C.~Yu.
\newblock Near neighbor join.
\newblock In {\em ICDE}, pages 1120--1131, 2014.

\bibitem{DBLP:conf/sigmod/KllapiSTI11}
H.~Kllapi, E.~Sitaridi, M.~M. Tsangaris, and Y.~E. Ioannidis.
\newblock Schedule optimization for data processing flows on the cloud.
\newblock In {\em SIGMOD Conference}, pages 289--300, 2011.

\bibitem{DBLP:journals/pvldb/LambFVTVDB12}
A.~Lamb et~al.
\newblock The vertica analytic database: C-store 7 years later.
\newblock {\em PVLDB}, 5(12):1790--1801, 2012.

\bibitem{DBLP:conf/icdcs/LeeLHWHZ11}
R.~Lee, T.~Luo, Y.~Huai, F.~Wang, Y.~He, and X.~Zhang.
\newblock Ysmart: Yet another sql-to-mapreduce translator.
\newblock In {\em {ICDCS}}, pages 25--36, 2011.

\bibitem{DBLP:journals/pvldb/LowGKBGH12}
Y.~Low, J.~Gonzalez, A.~Kyrola, D.~Bickson, C.~Guestrin, and J.~M. Hellerstein.
\newblock Distributed graphlab: A framework for machine learning in the cloud.
\newblock {\em PVLDB}, 5(8):716--727, 2012.

\bibitem{DBLP:journals/pvldb/MelnikGLRSTV10}
S.~Melnik, A.~Gubarev, J.~J. Long, G.~Romer, S.~Shivakumar, M.~Tolton, and
  T.~Vassilakis.
\newblock Dremel: Interactive analysis of web-scale datasets.
\newblock {\em {PVLDB}}, 3(1):330--339, 2010.

\bibitem{DBLP:reference/db/MorfoniosI09a}
K.~Morfonios and Y.~E. Ioannidis.
\newblock Snowflake schema.
\newblock In {\em Encyclopedia of Database Systems}, pages 2665--2666. 2009.

\bibitem{DBLP:conf/sigmod/OkcanR11}
A.~Okcan and M.~Riedewald.
\newblock Processing theta-joins using mapreduce.
\newblock In {\em SIGMOD Conference}, pages 949--960, 2011.

\bibitem{DBLP:conf/sigmod/OlstonRSKT08}
C.~Olston, B.~Reed, U.~Srivastava, R.~Kumar, and A.~Tomkins.
\newblock Pig latin: a not-so-foreign language for data processing.
\newblock In {\em {SIGMOD}}, pages 1099--1110, 2008.

\bibitem{DBLP:journals/sp/PikeDGQ05}
R.~Pike, S.~Dorward, R.~Griesemer, and S.~Quinlan.
\newblock Interpreting the data: Parallel analysis with {Sawzall}.
\newblock {\em Scientific Programming}, 13(4):277--298, 2005.

\bibitem{DBLP:conf/damon/PolychroniouR13}
O.~Polychroniou and K.~A. Ross.
\newblock High throughput heavy hitter aggregation for modern simd processors.
\newblock In {\em DaMoN}, page~6, 2013.

\bibitem{DBLP:conf/sigmod/SchaffnerJKKPFJ13}
J.~Schaffner, T.~Januschowski, M.~Kercher, T.~Kraska, H.~Plattner, M.~J.
  Franklin, and D.~Jacobs.
\newblock {RTP:} robust tenant placement for elastic in-memory database
  clusters.
\newblock In {\em {SIGMOD} Conference, New York, NY, USA, June 22-27}, pages
  773--784, 2013.

\bibitem{DBLP:conf/vldb/Simitsis03}
A.~Simitsis.
\newblock Modeling and managing etl processes.
\newblock In {\em VLDB PhD Workshop}, 2003.

\bibitem{DBLP:conf/icde/ThusooSJSCZALM10}
A.~Thusoo, J.~S. Sarma, N.~Jain, Z.~Shao, P.~Chakka, N.~Zhang, S.~Anthony,
  H.~Liu, and R.~Murthy.
\newblock Hive - a petabyte scale data warehouse using Hadoop.
\newblock In {\em {ICDE}, March 1-6, California, {USA}},
  pages 996--1005, 2010.

\bibitem{DBLP:conf/edbt/TinnefeldKGBRSP13}
C.~Tinnefeld, D.~Kossmann, M.~Grund, J.~Boese, F.~Renkes, V.~Sikka, and
  H.~Plattner.
\newblock Elastic online analytical processing on ramcloud.
\newblock In {\em Joint 2013 {EDBT/ICDT} Conferences, {EDBT} '13 Proceedings,
  Genoa, Italy, March 18-22, 2013}, pages 454--464, 2013.

\bibitem{DBLP:conf/icde/TsakalozosKSRPD11}
K.~Tsakalozos, H.~Kllapi, E.~Sitaridi, M.~Roussopoulos, D.~Paparas, and
  A.~Delis.
\newblock Flexible use of cloud resources through profit maximization and price
  discrimination.
\newblock In {\em {ICDE}, April 11-16, Hannover, Germany}, pages 75--86,
  2011.

\bibitem{trfd-icdcs10}
K.~Tsakalozos, M.~Roussopoulos, V.~Floros, and A.~Delis.
\newblock {Nefeli: Hint-based Execution of Workloads in Clouds }.
\newblock In {\em {ICDCS}}, 2010.

\bibitem{DBLP:journals/debu/TsangarisKKPPPSSI09}
M.~M. Tsangaris, G.~Kakaletris, H.~Kllapi, G.~Papanikos, F.~Pentaris,
  P.~Polydoras, E.~Sitaridi, V.~Stoumpos, and Y.~E. Ioannidis.
\newblock Dataflow processing and optimization on grid and cloud
  infrastructures.
\newblock {\em {IEEE} Data Eng. Bull.}, 32(1):67--74, 2009.

\bibitem{DBLP:journals/toms/Vitter85}
J.~S. Vitter.
\newblock Random sampling with a reservoir.
\newblock {\em ACM Trans. Math. Softw.}, 11(1):37--57, 1985.

\bibitem{DBLP:conf/sigmod/XinRZFSS13}
R.~S. Xin, J.~Rosen, M.~Zaharia, M.~J. Franklin, S.~Shenker, and I.~Stoica.
\newblock Shark: {SQL} and rich analytics at scale.
\newblock In {\em {SIGMOD} Conference, New York, NY, USA, June 22-27}, pages
  13--24, 2013.

\bibitem{xiong2014admission}
P.~Xiong et~al.
\newblock Admission control in cloud databases under service level agreements,
  July~1 2014.
\newblock US Patent 8,768,875.

\end{thebibliography}
